\documentclass[12pt,a4paper]{article}
\usepackage{amssymb}
\usepackage{amsmath}
\usepackage{float}
\usepackage{empheq}
\begin{document}

\begin{center}
{\Large \bf Fermions in five-dimensional brane world models}

\vspace{4mm}

Mikhail N.~Smolyakov\\
\vspace{0.5cm} Skobeltsyn Institute of Nuclear Physics, Lomonosov
Moscow State University,
\\ 119991, Moscow, Russia
\end{center}

\begin{abstract}
In the present paper the fermion fields, living in the background of five-dimensional warped brane world models with compact extra dimension, are thoroughly examined. The Kaluza-Klein decomposition and isolation of the physical degrees of freedom is performed for those five-dimensional fermion field Lagrangians, which admit such a decomposition to be performed in a mathematically consistent way and provide a physically reasonable four-dimensional effective theory. It is also shown that for the majority of five-dimensional fermion field Lagrangians there are no (at least rather obvious) ways to perform the Kaluza-Klein decomposition consistently. Moreover, in these cases one may expect the appearance of various pathologies in the four-dimensional effective theory. Among the cases, for which the Kaluza-Klein decomposition can be performed in a mathematically consistent way, the case, which reproduces the Standard Model by the zero Kaluza-Klein modes most closely regardless of the size of the extra dimension, is examined in detail in the background of the Randall-Sundrum model.
\end{abstract}

\section{Introduction}
In the original formulation of modern models with extra dimensions \cite{ArkaniHamed:1998rs,Randall:1999ee}, the Standard Model fields were supposed to be located on a brane. However, very soon the idea of theories with universal extra dimensions, in which all fields can propagate in extra dimensions, was proposed \cite{Appelquist:2000nn}. In this case the fields posses their own towers of Kaluza-Klein modes, representing the physical degrees of freedom of the four-dimensional effective theory; the zero modes are supposed to be the Standard Model fields. By now, there exist many papers describing different scenarios with universal extra dimensions and their phenomenological consequences.

In order to perform the Kaluza-Klein decomposition consistently in a particular case, it would be well to have a detailed procedure of Kaluza-Klein decomposition in the general cases, at least for frequently used background metrics. Such procedures for the scalar and gauge fields are rather obvious and universal and can be found in the literature. As for fermion fields, in the simplest case of one flat compact extra dimension all the key ingredients, which are necessary to perform the Kaluza-Klein decomposition in a mathematically consistent way are also known and were presented, for example, in \cite{Macesanu}. In the model with infinite flat extra dimension, admitting localization of the massive fermion zero mode via the Rubakov-Shaposhnikov mechanism \cite{RS}, the corresponding decomposition was discussed in detail in \cite{Smolyakov:2011hv}. However, I have failed to find such an analysis in the general case for another widely used branch of five-dimensional brane world models --- models with compact extra dimension and non-factorizable metric of the Randall-Sundrum type \cite{Randall:1999ee} (warped brane world models) in the case when the zero mode is supposed to have a nonzero mass (for example, generated via the Higgs mechanism). Usually, in this case the perturbation theory is used to describe the effective four-dimensional theory, treating the interaction with the Higgs field as a correction.

In the present paper an attempt is made to find a consistent method for performing the Kaluza-Klein decomposition of fermion fields, living in the bulk of five-dimensional warped brane world models with non-factorizable metric of the standard form, in a rather model-independent way. The term providing the zero mode fermion mass (which can originate, for example, from the interaction with the Higgs field) is not supposed to be a perturbation and is taken into account from the very beginning.

In the previous paper on this subject \cite{Smolyakov:2015zsa} it was shown that in the general case there may exist pathologies in the fermion sector of warped five-dimensional brane-world models. The more detailed analysis, which will be presented below, shows that the Kaluza-Klein decomposition for fermions can be performed consistently only for several five-dimensional fermion field Lagrangians, some of which admit a localization of the zero mode, but demand special choices of the mass generating term (i.e., the form of the Higgs field profile in the extra dimension) or the localizing term, whereas the other admit any form of the mass generating term, but forbid a localization of the zero mode (for the smooth mass generating and localizing terms these five-dimensional Lagrangians reproduce those found earlier in \cite{Smolyakov:2015zsa}). All these cases provide second-order differential equations of motion for the components of the fermion fields (or of their linear combinations) in the whole five-dimensional space-time or in the whole five-dimensional space-time except the fixed points of the orbifold (where the branes are located). An interesting observation is that the procedures of Kaluza-Klein decomposition, which will be presented in detail below, differ considerably for different cases. As for the general case, I have failed to find an obvious way to perform the Kaluza-Klein decomposition, even for particular choices of the parameters, because of the appearance of fourth-order differential equations of motion for the fields. The latter seems impossible to circumvent, at least by the standard diagonalization procedures. Moreover, since higher-derivative theories are known to contain pathologies (this issue will be discussed in the text), one may expect the appearance of analogous pathologies in the case under consideration too.

A special choice of the five-dimensional Lagrangian was examined with relation to a possibility to reproduce the ordinary four-dimensional Standard Model by the zero Kaluza-Klein modes (of the fermion, gauge and Higgs fields) most closely regardless of the value of the five-dimensional energy scale and without taking into account the higher Kaluza-Klein modes. As a particular background, the Randall-Sundrum model \cite{Randall:1999ee} was considered. It was shown that for appropriate values of the parameters of the model deviations from the Standard Model appear only in the coupling constants of the Higgs boson to fermions and in the Higgs boson self-coupling constants, the corresponding corrections are presented in an explicit form.

The paper is organized as follows. In Section~2 we demonstrate a simple introductory example of the consistent Kaluza-Klein decomposition procedure for the model with flat compact extra dimension. In Section~3 we examine the general case of five-dimensional warped brane world models and find the five-dimensional Lagrangians for which the Kaluza-Klein decomposition can be performed in a mathematically consistent way. We also discuss the problems, which can arise in the general case. In Section~4 the Kaluza-Klein decomposition is performed in general for three different cases, for which this can be done analytically in a mathematically consistent way. In Section~5 a model, based on the Randall-Sundrum background metric, is considered as a possible candidate that allows one to reproduce the Standard Model by the zero mode sector with a minimal possible set of restrictions coming from the effective theory composed of these zero modes. And finally, in Section~6 we briefly discuss the obtained results. Auxiliary calculations are collected in the Appendices.

\section{Flat five-dimensional space-time}
As a simple introductory example let us take a flat five-dimensional space-time with the coordinates
$x^{M}=\{x^{\mu},z\}$, $M=0,1,2,3,5$. The compact extra dimension
is supposed to form the orbifold $S^1/Z_2$, which can be
represented as the circle with the coordinate $-L\le z\le L$ and
the points $-z$ and $z$ identified. In what follows, we will use
the notation $x$ for the coordinates $x^{\mu}$.

It is well known that there is no chirality in five-dimensional space-time. Thus, in order to obtain a nonzero
mass term for the zero Kaluza-Klein fermion mode, it is necessary to take two five-dimensional spinor
fields (see, for example, \cite{Macesanu,DRT,CG}) satisfying the orbifold
symmetry conditions
\begin{eqnarray}\label{sym01}
\Psi_{1}(x,-z)=\gamma^{5}\Psi_{1}(x,z),\\ \label{sym02}
\Psi_{2}(x,-z)=-\gamma^{5}\Psi_{2}(x,z).
\end{eqnarray}
Thus, as an example we consider a  model with the action of the form
\begin{eqnarray}\label{fermactUED}
S=\int d^{4}xdz\Bigl(i\bar\Psi_{1}\Gamma^{M}\partial_{M}\Psi_{1}+i\bar\Psi_{2}\Gamma^{M}\partial_{M}\Psi_{2}-M\left(\bar\Psi_{1}\Psi_{2}+\bar\Psi_{2}\Psi_{1}\right)\Bigr),
\end{eqnarray}
where $\Gamma^{\mu}=\gamma^{\mu}$, $\Gamma^{5}=i\gamma^{5}$, $M>0$. From here and below we will use the chiral representation of the gamma matrices
\begin{equation}
\gamma^{0}=\begin{pmatrix}
0&I\\
I&0\\
\end{pmatrix},\qquad \gamma^{i}=\begin{pmatrix}
0&\sigma^{i}\\
-\sigma^{i}&0\\
\end{pmatrix},
\end{equation}
in which $\gamma^5$ is diagonal and has the form
\begin{equation}
\gamma^{5}=\begin{pmatrix}
I&0\\
0&-I
\end{pmatrix}.
\end{equation}

The orbifold symmetry conditions (\ref{sym01}), (\ref{sym02}), the geometric structure of the extra dimension and the form of action (\ref{fermactUED}) suggest the standard Fourier decomposition for the five-dimensional fermion fields $\Psi_{1}$ and $\Psi_{2}$, i.e., these fields can be decomposed into Kaluza-Klein modes as (see
\cite{Macesanu})
\begin{eqnarray}\label{subst1}
\Psi_{1}(x,z)=\frac{1}{\sqrt{2L}}\psi_{L}(x)+\frac{1}{\sqrt{L}}\sum\limits_{n=1}^{\infty}\left(\cos\left(\frac{\pi
n}{L}z\right)\psi^{n}_{L}(x)-\sin\left(\frac{\pi
n}{L}z\right)\hat\psi^{n}_{R}(x)\right),\\ \label{subst2}
\Psi_{2}(x,z)=\frac{1}{\sqrt{2L}}\psi_{R}(x)+\frac{1}{\sqrt{L}}\sum\limits_{n=1}^{\infty}\left(\cos\left(\frac{\pi
n}{L}z\right)\psi^{n}_{R}(x)+\sin\left(\frac{\pi
n}{L}z\right)\hat\psi^{n}_{L}(x)\right),
\end{eqnarray}
where $\psi_{L}(x)=\gamma^{5}\psi_{L}(x)$,
$\psi_{R}(x)=-\gamma^{5}\psi_{R}(x)$,
$\psi^{n}_{L}(x)=\gamma^{5}\psi^{n}_{L}(x)$,
$\psi^{n}_{R}(x)=-\gamma^{5}\psi^{n}_{R}(x)$,
$\hat\psi^{n}_{L}(x)=\gamma^{5}\hat\psi^{n}_{L}(x)$,
$\hat\psi^{n}_{R}(x)=-\gamma^{5}\hat\psi^{n}_{R}(x)$. Note that the components of the five-dimensional fields (\ref{subst1}) and (\ref{subst2}) satisfy the five-dimensional equations
\begin{equation}\label{KGsec2}
\Box\Psi_{1}-\Psi_{1}''+M^{2}\Psi_{1}=0,\qquad \Box\Psi_{2}-\Psi_{2}''+M^{2}\Psi_{2}=0
\end{equation}
with
\begin{equation}
\Box\psi_{L,R}+M^{2}\psi_{L,R}=0,\qquad \Box\psi^{n}_{L,R}+m_{n}^{2}\psi^{n}_{L,R}=0,\qquad \Box\hat\psi^{n}_{L,R}+m_{n}^{2}\hat\psi^{n}_{L,R}=0,
\end{equation}
where $m_{n}=\sqrt{\frac{\pi^2n^2}{L^2}+M^2}$. Equations (\ref{KGsec2}) follow directly from the five-dimensional Dirac equations, originating from (\ref{fermactUED}). In fact, it is equations (\ref{KGsec2}) that suggest the correct separation of variables, determine the form of the wave functions in (\ref{subst1}), (\ref{subst2}), define the mass spectrum and guarantee that (\ref{subst1}) and (\ref{subst2}) indeed describe the complete set of possible physical degrees of freedom of the theory. It is clear that in other models the equations analogous to (\ref{KGsec2}) may have different form, leading to wave functions different from the components of the Fourier series.

Substituting (\ref{subst1}) and (\ref{subst2}) into (\ref{fermactUED}) and
integrating over the coordinate $z$ of the extra dimension, we
arrive at
\begin{eqnarray}\nonumber
S_{eff}=\int d^{4}x
\Biggl(i\bar\psi\gamma^{\mu}\partial_{\mu}\psi-M\bar\psi\psi+\sum\limits_{n=1}^{\infty}\left.\Bigl(
i\bar\psi^{n}\gamma^{\mu}\partial_{\mu}\psi^{n}\right.\\
\label{act4stand} \left.+i\bar{\hat\psi}^{n}\gamma^{\mu}\partial_{\mu}\hat\psi^{n}-\frac{\pi n}{L}(\bar{\hat\psi}^{n}\psi^{n}+\bar\psi^{n}\hat\psi^{n})-
M(\bar\psi^{n}\psi^{n}-\bar{\hat\psi}^{n}\hat\psi^{n})\right)\Biggr)
\end{eqnarray}
with
\begin{eqnarray}
\psi(x)=\psi_{L}(x)+\psi_{R}(x),\\
\psi^n(x)=\psi_{L}^n(x)+\psi_{R}^n(x),\\
\hat\psi^n(x)=\hat\psi_{L}^n(x)+\hat\psi_{R}^n(x).
\end{eqnarray}
We see that the mass matrix is non-diagonal \cite{Macesanu}. In order to bring it
to the diagonal form, we use the transformations \cite{Smolyakov:2012ud}
\begin{eqnarray}\label{rotation01}
\psi^{n}(x)&=&\psi_{1}^{n}(x)\cos(\theta_{n})+\psi_{2}^{n}(x)\sin(\theta_{n}),\\ \label{rotation02}
\hat\psi^{n}(x)&=&\psi_{1}^{n}(x)\sin(\theta_{n})-\psi_{2}^{n}(x)\cos(\theta_{n})
\end{eqnarray}
where $n\ge 1$ and
\begin{equation}\label{tanthetan}
\tan(2\theta_{n})=\frac{\pi n}{ML}
\end{equation}
and obtain
\begin{eqnarray}\nonumber
S_{eff}=\int d^{4}x
\Biggl(i\bar\psi\gamma^{\mu}\partial_{\mu}\psi-M\bar\psi\psi\\ \label{act4unc}+\sum\limits_{n=1}^{\infty}\Bigl(
i\bar\psi_{1}^{n}\gamma^{\mu}\partial_{\mu}\psi_{1}^{n}+i\bar\psi_{2}^{n}\gamma^{\mu}\partial_{\mu}\psi_{2}^{n}-
m_{n}(\bar{\psi}_{1}^{n}\psi_{1}^{n}-\bar{\psi}_{2}^{n}\psi_{2}^{n})\Bigr)\Biggr),
\end{eqnarray}
where $m_{n}=\sqrt{\frac{\pi^2n^2}{L^2}+M^2}$. We see that the
mass terms of the fields $\psi_{2}^{n}$ have the unconventional sign.
But with the help of the standard redefinition $\psi_{2}^{n}\to
\gamma^{5}\psi_{2}^{n}$, we can bring action (\ref{act4unc}) to the standard form
\begin{eqnarray}
S_{eff}=\int d^{4}x
\left(i\bar\psi\gamma^{\mu}\partial_{\mu}\psi-M\bar\psi\psi+\sum\limits_{n=1}^{\infty}\sum\limits_{i=1}^{2}\Bigl(
i\bar\psi_{i}^{n}\gamma^{\mu}\partial_{\mu}\psi_{i}^{n}-m_{n}\bar{\psi}_{i}^{n}\psi_{i}^{n}\Bigr)\right).
\end{eqnarray}
This four-dimensional effective action describes the Dirac fermion with mass $M$, which is the lowest Kaluza-Klein mode, and a pair of two Dirac fermions with the same four-dimensional mass $m_{n}$ at each Kaluza-Klein level with $n\ge 1$.

We see that the Kaluza-Klein decomposition in this case is not complicated. However, it is so mainly because in the case under consideration we can use the Fourier decomposition for the fields from the very beginning. It is not so in the general case, for which the Fourier decomposition clearly does not provide the appropriate set of eigenfunctions and eigenvalues, especially in the curved background of brane world models. This problem will be discussed in the next sections.

\section{Warped five-dimensional space-time}
\subsection{Equations of motion}
Again, let us take a five-dimensional space-time with the compact extra dimension forming the orbifold $S^1/Z_2$ with the coordinate $-L\le z\le L$ and the points $-z$ and $z$ identified. Let us consider the following form of the background metric, which is standard in five-dimensional brane world models:
\begin{equation}\label{backgmetric}
ds^2=e^{2\sigma(z)}\eta_{\mu\nu}dx^{\mu}dx^{\nu}-dz^2,
\end{equation}
where $\sigma(-z)=\sigma(z)$. We do not specify the explicit form of the solution for $\sigma(z)$.

As an example, we consider a model with the action of the general form
\begin{eqnarray}\nonumber
S=\int d^{4}xdz\sqrt{g}\left(E_{N}^{M}i\bar\Psi_{1}\Gamma^{N}\nabla_{M}\Psi_{1}+E_{N}^{M}i\bar\Psi_{2}\Gamma^{N}\nabla_{M}\Psi_{2}\right.\\
\label{faction}\left.-F(z)\left(\bar\Psi_{1}\Psi_{1}-\bar\Psi_{2}\Psi_{2}\right)-G(z)\left(\bar\Psi_{2}\Psi_{1}+\bar\Psi_{1}\Psi_{2}\right)\right),
\end{eqnarray}
where $M,N=0,1,2,3,5$, $\Gamma^{\mu}=\gamma^{\mu}$, $\Gamma^{5}=i\gamma^{5}$, $\nabla_{M}$ is the covariant derivative containing the spin connection, $E_{N}^{M}$ is the vielbein. The fields are also supposed to satisfy the orbifold symmetry conditions
\begin{eqnarray}\label{sym1}
\Psi_{1}(x,-z)=\gamma^{5}\Psi_{1}(x,z),\\ \label{sym2}
\Psi_{2}(x,-z)=-\gamma^{5}\Psi_{2}(x,z),
\end{eqnarray}
the functions $F(z)$ and $G(z)$ are such that $F(-z)=-F(z)$ and $G(-z)=G(z)$. Again, due to the absence of chirality in five-dimensional space-time, we have to take two five-dimensional spinor fields $\Psi_{1}$ and $\Psi_{2}$ \cite{Macesanu,DRT,CG}.

As we will see below, the function $F(z)$ is responsible for the localization of the zero Kaluza-Klein mode, whereas the function $G(z)$ provides its nonzero mass. Contrary to the case of the previous section, here the background metric depends on the coordinate of the extra dimension $z$, so we also assume that the function $G(z)$ (which, of course, can be connected with the vacuum solution of some five-dimensional Higgs-like field) can also depend on the coordinate of the extra dimension.

For the case of metric (\ref{backgmetric}), action (\ref{faction}) can be rewritten in the form (see, for example, \cite{CG,fRS} for the
explicit form of the vielbein and spin connections)
\begin{eqnarray}\nonumber
S=\int d^{4}xdze^{4\sigma}\left(e^{-\sigma}i\bar\Psi_{1}\gamma^{\mu}\partial_{\mu}\Psi_{1}-\bar\Psi_{1}\gamma^{5}\left(\partial_{5}+2\sigma'\right)\Psi_{1}
\right.\\ \nonumber\left.
+e^{-\sigma}i\bar\Psi_{2}\gamma^{\mu}\partial_{\mu}\Psi_{2}-\bar\Psi_{2}\gamma^{5}\left(\partial_{5}+2\sigma'\right)\Psi_{2}\right.\\ \label{feffact}\left.
-F(z)\left(\bar\Psi_{1}\Psi_{1}-\bar\Psi_{2}\Psi_{2}\right)-G(z)\left(\bar\Psi_{2}\Psi_{1}+\bar\Psi_{1}\Psi_{2}\right)\right).
\end{eqnarray}
The equations of motion for the fields $\Psi_{1}$ and $\Psi_{2}$ take the form
\begin{eqnarray}\label{feqs1}
e^{-\sigma}i\gamma^{\mu}\partial_{\mu}\Psi_{1}-\gamma^{5}\left(\partial_{5}+2\sigma'\right)\Psi_{1}
-F\Psi_{1}-G\Psi_{2}=0,\\ \label{feqs2}
e^{-\sigma}i\gamma^{\mu}\partial_{\mu}\Psi_{2}-\gamma^{5}\left(\partial_{5}+2\sigma'\right)\Psi_{2}
+F\Psi_{2}-G\Psi_{1}=0.
\end{eqnarray}
It is clear the if $G(z)\equiv 0$, then there always exists the solution
\begin{eqnarray}
\Psi_{1}=C_{f}\exp\left[-\int\limits_{0}^{z}F(y)dy-2\sigma(z)\right]\psi_{L}(x),\quad i\gamma^{\mu}\partial_{\mu}\psi_{L}=0,&& \gamma^{5}\psi_{L}=\psi_{L},\\
\Psi_{2}=C_{f}\exp\left[-\int\limits_{0}^{z}F(y)dy-2\sigma(z)\right]\psi_{R}(x),\quad i\gamma^{\mu}\partial_{\mu}\psi_{R}=0,&& \gamma^{5}\psi_{R}=-\psi_{R},
\end{eqnarray}
where $C_{f}$ is a normalization constant. This solution describes two massless four-dimensional chiral fermions, clearly indicating that the term with $G(z)\not\equiv 0$ indeed provides a mass of the zero mode four-component fermion.

From (\ref{feqs1}) and (\ref{feqs2}) it is not difficult to obtain the second-order differential equations for the components of the fields $\Psi_{1}$ and $\Psi_{2}$:
\begin{eqnarray}\nonumber
-e^{-2\sigma}\Box\Psi_{1}+\Psi_{1}''+5\sigma'\Psi_{1}'+(6{\sigma'}^2+2\sigma'')\Psi_{1}-(G^{2}+F^{2}-e^{-\sigma}(e^{\sigma}F)'\gamma^{5})\Psi_{1}\\ \label{KGfeqs1} +
e^{-\sigma}(e^{\sigma}G)'\gamma^{5}\Psi_{2}=0,\\
\nonumber
-e^{-2\sigma}\Box\Psi_{2}+\Psi_{2}''+5\sigma'\Psi_{2}'+(6{\sigma'}^2+2\sigma'')\Psi_{2}-(G^{2}+F^{2}+e^{-\sigma}(e^{\sigma}F)'\gamma^{5})\Psi_{2}\\ \label{KGfeqs2}
+e^{-\sigma}(e^{\sigma}G)'\gamma^{5}\Psi_{1}=0.
\end{eqnarray}
These equations are not supposed to be used for $G(z)\sim\delta(z)$, this case will be considered separately in Section~4.

An important remark is in order here. In the case of an ordinary free four-dimensional fermion, satisfying the Dirac equation, each component of the fermion field satisfies the second-order Klein-Gordon differential equation. Of course, not all the components of this field are independent: one can choose, for example, $\psi_{L}$ as an independent part of the field, satisfying the Klein-Gordon equation, whereas $\psi_{R}$ can be expressed through $\psi_{L}$ with the help of the initial Dirac equation. Since in our case the fields $\Psi_{1}$ and $\Psi_{2}$ interact only with the background metric and with the ``fields'' $G(z)$ and $F(z)$, which correspond to the vacuum configuration, $\Psi_{1}$ and $\Psi_{2}$ should be considered as free fields as well. In this case one can try to choose, say, $\Psi_1$ (or some linear combination of the fields $\Psi_{1}$ and $\Psi_{2}$) as an independent field and express $\Psi_2$ through $\Psi_{1}$ with the help of equation (\ref{feqs1}) (or with the help of some combination of equations (\ref{feqs1}) and (\ref{feqs2}) for the linear combination of the fields) in full analogy with the four-dimensional case. One may expect that in a consistent theory the equations of motion for the components of the fields $\Psi_{1}$ and $\Psi_{2}$ (or of their linear combinations) should not contain any pathologies, otherwise such pathologies would probably arise in the resulting effective theory. So, we expect that the corresponding equations of motion should be five-dimensional second-order differential equations, which contain the derivatives in the four-dimensional coordinates only in the form $\Box=\eta^{\mu\nu}\partial_{\mu}\partial_{\nu}$ and provide a correct mass spectrum for the Kaluza-Klein modes. Moreover, as will be shown in Section~4, such second-order equations of motion indeed allow one to perform the Kaluza-Klein decomposition consistently. However, such equations can be obtained not in all cases.

\subsection{Decoupling the equations of motion}
We see that equations (\ref{KGfeqs1}), (\ref{KGfeqs2}) remain coupled, whereas equations for {\em each} component of the fields $\Psi_{1}$ and $\Psi_{2}$ separately can be obtained only for the special choice
\begin{equation}\label{finetune}
G(z)\equiv Me^{-\sigma(z)},
\end{equation}
where $M$ is a constant, which provides $\left(e^{\sigma}G\right)'\equiv 0$. In this case we get for (\ref{KGfeqs1}) and (\ref{KGfeqs2})
\begin{eqnarray}\label{KGfeqs3}
-e^{-2\sigma}(\Box\Psi_{1}+M^2\Psi_{1})+\Psi_{1}''+5\sigma'\Psi_{1}'+(6{\sigma'}^2+2\sigma'')\Psi_{1}-(F^{2}-e^{-\sigma}(e^{\sigma}F)'\gamma^{5})\Psi_{1}=0,\\
\label{KGfeqs4}
-e^{-2\sigma}(\Box\Psi_{2}+M^2\Psi_{2})+\Psi_{2}''+5\sigma'\Psi_{2}'+(6{\sigma'}^2+2\sigma'')\Psi_{2}-(F^{2}+e^{-\sigma}(e^{\sigma}F)'\gamma^{5})\Psi_{2}=0.
\end{eqnarray}
From these equations it is easy to get the solution for the lowest mode, which takes the form
\begin{eqnarray}\label{zm1}
\Psi_{1}=C_{f}\exp\left[-\int\limits_{0}^{z}F(y)dy-2\sigma(z)\right]\psi_{L}(x),\quad
i\gamma^{\mu}\partial_{\mu}\psi_{L}-M\psi_{R}=0,&&
\gamma^{5}\psi_{L}=\psi_{L},\\ \label{zm2}
\Psi_{2}=C_{f}\exp\left[-\int\limits_{0}^{z}F(y)dy-2\sigma(z)\right]\psi_{R}(x),\quad
i\gamma^{\mu}\partial_{\mu}\psi_{R}- M\psi_{L}=0,&&
\gamma^{5}\psi_{R}=-\psi_{R},
\end{eqnarray}
where again $C_{f}$ is a normalization constant. The fields $\psi_{L}$ and $\psi_{R}$ are localized in the vicinity of the same point in the extra dimension because they have the same wave function; taken together they make up a four-dimensional Dirac fermion with mass $M$. It is not difficult to show that solution (\ref{zm1}) and (\ref{zm2}) indeed stands for the lowest mode, see the proof in \cite{Smolyakov:2015zsa}.

What can we learn from equations (\ref{zm1}), (\ref{zm2})? First, it is the term with the five-dimensional mass $M$ that provides the mass of the zero Kaluza-Klein mode. Second, the wave functions of the left and right component of the field $\psi$ are exactly the same. The latter means that, depending on the form of the function $F(z)$, the whole zero mode (composed of the left and right two-component spinors) can be localized at any point of the orbifold (see, for example, \cite{RS,DRT,AAGS}). Very often in brane world models the function $F(z)$ is taken in the form $F(z)=Q\,\textrm{sign}(z)$, where
the value of the constant $Q$ defines at which brane the fermion zero mode is localized and what is the width of its wave function
\cite{DRT,Grossman:1999ra,Gherghetta:2000qt,Huber:2000ie}.

We will not perform the complete Kaluza-Klein decomposition here (the special case (\ref{finetune}) will be discussed in detail in Section~4) and proceed to the more general case $(e^{\sigma}G)'\not\equiv 0$.

In order to decouple equations (\ref{KGfeqs1}) and (\ref{KGfeqs2}) for $(e^{\sigma}G)'\not\equiv 0$, it is convenient first to consider the left-handed parts of the spinor fields $\Psi_{1}^{L}=\gamma^{5}\Psi_{1}^{L}$, $\Psi_{2}^{L}=\gamma^{5}\Psi_{2}^{L}$. Following \cite{Smolyakov:2015zsa}, equations (\ref{KGfeqs1}) and (\ref{KGfeqs2}) for $\Psi_{1}^{L}$ and $\Psi_{2}^{L}$ can be rewritten in the operator form as
\begin{equation}\label{eqsoper}
\begin{pmatrix}
\hat L& 0\\
0& \hat L\\
\end{pmatrix}\begin{pmatrix}
\Psi_{1}^{L}\\
\Psi_{2}^{L}\\
\end{pmatrix}+{\hat\Lambda}_{L}\begin{pmatrix}
\Psi_{1}^{L}\\
\Psi_{2}^{L}\\
\end{pmatrix}=0,
\end{equation}
where the operator $\hat L$ is defined as
\begin{equation}\label{Loperatordef}
\hat L=-e^{-2\sigma}\Box+\partial_{5}^{2}+5\sigma'\partial_{5}+6{\sigma'}^2+2\sigma''-G^{2}-F^{2}
\end{equation}
and the matrix $\hat\Lambda_{L}$ contains only the functions depending on the coordinate $z$ and looks like
\begin{equation}\label{HatLambdadef}
{\hat\Lambda}_{L} =\begin{pmatrix}
e^{-\sigma}(e^{\sigma}F)'& e^{-\sigma}(e^{\sigma}G)'\\
e^{-\sigma}(e^{\sigma}G)'&-e^{-\sigma}(e^{\sigma}F)'\\
\end{pmatrix}.
\end{equation}
The form of equation (\ref{eqsoper}) implies that the decoupling of the equations of motion for the components of the fermion fields is equivalent to
the diagonalization of the matrix ${\hat\Lambda}_{L}$. This matrix is symmetric, so it can be diagonalized in the standard way with the help of a rotation matrix \begin{equation}
{\hat R}=\begin{pmatrix}
\cos \Theta& -\sin \Theta\\
\sin \Theta&\cos \Theta\\
\end{pmatrix}, \qquad {\hat R}^T{\hat\Lambda}_{L}{\hat R}=\textrm{diag}(\lambda_{1},\lambda_{2}).
\end{equation}
The rotation angle can be easily found by the standard procedure and takes the form
\begin{equation}
\cot(2\Theta)=\frac{(e^{\sigma}F)'}{(e^{\sigma}G)'}.
\end{equation}
It is clear that in the general case the rotation angle $\Theta$ depends on the coordinate of the extra dimension $z$. On the other hand, the rotation angle $\Theta$ should not depend on the coordinate of the extra dimension, otherwise the rotation matrix $\hat R$ would not pass through the operator $\hat L$, which contains derivatives in $z$. The obvious exception is
\begin{equation}\label{Feqzero}
F(z)\equiv 0,
\end{equation}
i.e., we turn off the localization mechanism.\footnote{In principle, one can consider a more general form of the ``localizing term'', i.e., $F_{1}(z)\bar\Psi_{1}\Psi_{1}+F_{2}(z)\bar\Psi_{2}\Psi_{2}$ with $F_{1,2}(-z)=-F_{1,2}(z)$ instead of $F(z)\left(\bar\Psi_{1}\Psi_{1}-\bar\Psi_{2}\Psi_{2}\right)$ in (\ref{faction}). In this case one can obtain $F_{1}(z)\equiv F_{2}(z)$ instead of (\ref{Feqzero}) \cite{Smolyakov:2015zsa}. However, in the case $G(z)\equiv 0$ the zero-mode massless fields $\psi_{L}$ and $\psi_{R}$, which are supposed to make up a four-dimensional Dirac fermion for $G(z)\not\equiv 0$, appear to have different wave functions in the extra dimension and, for some choices of $F_{1}(z)\equiv F_{2}(z)$, to be localized at the opposite points of the extra dimension, which looks rather unnatural. So, I use the standard choice $F_{2}(z)\equiv -F_{1}(z)$ \cite{DRT,CG,AAGS} from the very beginning.} In this case we get
\begin{equation}\label{thetaang}
\Theta=\frac{\pi}{4}.
\end{equation}
An analogous reasoning can be applied to $\Psi_{1}^{R}$ and $\Psi_{2}^{R}$ using the corresponding matrix $\hat\Lambda_{R}$. In fact, $\Theta=\frac{\pi}{4}$ corresponds to the combinations $\Psi_{1}+\Psi_{2}$ and $\Psi_{1}-\Psi_{2}$ of the fields $\Psi_{1}$, $\Psi_{2}$. Indeed, for $F(z)\equiv 0$ we can simply add and subtract equations (\ref{KGfeqs1}) and (\ref{KGfeqs2}) to get independent second-order differential equations for the components of the combinations $\Psi_{1}+\Psi_{2}$ and $\Psi_{1}-\Psi_{2}$.

It seems that there are no other possibilities to diagonalize the matrix $\hat\Lambda_{L}$ (and, correspondingly, the analogous matrix $\hat\Lambda_{R}$) using one and the same $\Theta$ in the whole space-time except if $G(z)\equiv Me^{-\sigma(z)}$ or $F(z)\equiv 0$ \cite{Smolyakov:2015zsa} (recall the symmetry properties of $F(z)$ and $G(z)$). However, if we consider generalized functions (like $\textrm{sign}(z)$ or $\delta(z)$), there is still a possibility to find other cases for which the matrices $\hat\Lambda_{L,R}$ can be diagonalized. Indeed, the case $F(z)\equiv 0$ allows one to get independent second-order differential equations for the whole extra dimension. But when one considers generalized functions, it is possible to get second-order differential equations of motion everywhere except the points $z=0$, $z=L$ of the extra dimension or even everywhere except a single point of the extra dimension. Note that generalized functions should not be considered as approximations of some continuous and differentiable functions, they should be taken as pure generalized functions in their mathematical sense. Namely, there are the following possibilities:
\begin{eqnarray}\label{partic1}
G(z)\equiv\gamma\,\textrm{sign}(z)F(z)+Me^{-\sigma},\quad\textrm{any}\quad F(z);\\ \label{partic2}
F(z)\equiv\gamma\,\textrm{sign}(z)e^{-\sigma},\quad\textrm{any}\quad G(z);\\ \label{partic3}
G(z)\equiv K_{1}\,\delta(z)+K_{2}\,\delta(z-L)+Me^{-\sigma},\quad\textrm{any}\quad F(z);
\end{eqnarray}
where $\gamma$, $K_{1}$ and $K_{2}$ are constants. For example, for $F(z)\sim \textrm{sign}(z)$ in (\ref{partic1}) and with $M=0$ we get $G(z)\sim\textrm{const}$. In the first case (\ref{partic1}) one can diagonalize the matrices $\hat\Lambda_{L,R}$ in the regions $-L<z<0$ and $0<z<L$ separately, then match the corresponding solutions at the points $z=0$ and $z=L$ using the boundary conditions following from equations (\ref{feqs1}), (\ref{feqs2}). In the second case (\ref{partic2}) one can use the second-order differential equations for the combinations $\Psi_{1}+\Psi_{2}$ and $\Psi_{1}-\Psi_{2}$ in the regions $-L<z<0$ and $0<z<L$, then also match the corresponding solutions at the points $z=0$ and $z=L$ using the boundary conditions following from equations (\ref{feqs1}), (\ref{feqs2}). And in the third case (\ref{partic3}) one can use equations (\ref{KGfeqs3}), (\ref{KGfeqs4}) everywhere except the points $z=0$, $z=L$ (or only except the point $z=L$ if $K_{1}=0$) and then use the matching conditions at $z=0$, $z=L$ (or only at the point $z=L$ if $K_{1}=0$) following from (\ref{feqs1}), (\ref{feqs2}).

All the special cases, presented above, will be discussed in more detail in Section~4, including the consistent procedures of the Kaluza-Klein decomposition for some of them. Now let us turn to examining the most general case, for which the second-order equations of motion can not be obtained, at least in a simple way.

\subsection{The general case: higher derivatives}
To examine in more detail the general case, in which the functions $F(z)$ and $G(z)$ do not satisfy the conditions (\ref{finetune}), (\ref{Feqzero}) or (\ref{partic1})--(\ref{partic3}), let us again, for simplicity, consider the left parts $\Psi_{1}^{L}$, $\Psi_{2}^{L}$ of the spinor fields and represent equations (\ref{KGfeqs1}) and (\ref{KGfeqs2}) as
\begin{eqnarray}\label{KGfeqs1L}
{\hat L}\Psi_{1}^{L}+e^{-\sigma}(e^{\sigma}F)'\Psi_{1}^{L}+e^{-\sigma}(e^{\sigma}G)'\Psi_{2}^{L}=0,\\
\label{KGfeqs2L}
{\hat L}\Psi_{2}^{L}-e^{-\sigma}(e^{\sigma}F)'\Psi_{2}^{L}+e^{-\sigma}(e^{\sigma}G)'\Psi_{1}^{L}=0,
\end{eqnarray}
where the operator $\hat L$ is defined by (\ref{Loperatordef}). If the appropriate diagonalization of the matrices $\hat\Lambda_{L,R}$ in (\ref{HatLambdadef}) is impossible, then the only obvious way is to obtain separate equations for the fields $\Psi_{1}^{L}$, $\Psi_{2}^{L}$. For example, equation for $\Psi_{1}^{L}$ takes the form
\begin{equation}\label{hder1}
\left(\hat L-e^{-\sigma}(e^{\sigma}F)'\right)\frac{\hat L+e^{-\sigma}(e^{\sigma}F)'}{e^{-\sigma}(e^{\sigma}G)'}\Psi_{1}^{L}-e^{-\sigma}(e^{\sigma}G)'\Psi_{1}^{L}=0.
\end{equation}
For $\Psi_{2}^{L}$ we can get
\begin{equation}\label{hder2}
\left(\hat L+e^{-\sigma}(e^{\sigma}F)'\right)\frac{\hat L-e^{-\sigma}(e^{\sigma}F)'}{e^{-\sigma}(e^{\sigma}G)'}\Psi_{2}^{L}-e^{-\sigma}(e^{\sigma}G)'\Psi_{2}^{L}=0.
\end{equation}
Analogous equations can be obtained for $\Psi_{1}^{R}$, $\Psi_{2}^{R}$. Thus, we get fourth-order differential equations of motion (recall that the operator $\hat L$ contains second derivatives). I have failed to solve these equations even for particular cases, however, in the general case such equations describe {\em more degrees of freedom} than the second-order differential equations (for example, in some cases one may expect the appearance of two ``lowest'' modes with close but different four-dimensional masses) and, in principle, may contain serious pathologies. To demonstrate the appearance of such pathologies explicitly, let us consider a simple four-dimensional model.

Indeed, let us take a four-dimensional scalar field theory with the action
\begin{equation}\label{4Dscact}
S_{4}=\int d^{4}x\left(\frac{\alpha}{M^{2}}\left(\Box\phi\right)^{2}+\partial_{\mu}\phi\partial^{\mu}\phi-M^{2}\phi^2\right),
\end{equation}
where $\alpha$ is a dimensionless parameter. If $\alpha=0$, we have the standard theory describing the scalar field of mass $M$, possessing the Klein-Gordon equation of motion. But if $\alpha\ne 0$, the equation of motion for the scalar field is the fourth-order differential equation
\begin{equation}\label{4Dsceq}
\frac{\alpha}{M^{2}}\Box^{2}\phi-\Box\phi-M^{2}\phi=0,
\end{equation}
describing two scalar degrees of freedom with masses defined by
\begin{equation}\label{csmasses}
m_{1,2}^{2}=M^{2}\frac{-1\pm\sqrt{1+4\alpha}}{2\alpha}.
\end{equation}
Suppose that $|\alpha|\ll 1$. In this case
\begin{equation}\label{4Dperturbroots}
m_{1}^{2}\approx M^{2}-\alpha M^{2},\qquad m_{2}^{2}\approx-\frac{M^{2}}{\alpha}.
\end{equation}
We see that for $\alpha>0$ the second root describes a tachyonic mode, indicating the existence of a classical instability. It should be noted that for $|\alpha|\ll 1$ the first root can be obtained perturbatively in $\alpha$, whereas it is not so for the second root. A much more detailed analysis of the inapplicability of the perturbation theory for examining the fourth-order differential equations can be found in a nice review \cite{Woodard:2006nt}. Of course, in the simple example (\ref{4Dscact}) the ``pathological'' term $\frac{\alpha}{M^{2}}\left(\Box\phi\right)^{2}$ was introduced ``by hands'', whereas equation (\ref{hder1}) with higher derivatives arises in a different way. However, from the mathematical point of view both equations are of the same kind.

Analogous pathologies may arise in the case of fourth-order equations (\ref{hder1}) and (\ref{hder2}), even if the term with the function $G(z)$ looks like a perturbation. The appearance of the fourth-order equations of motion for the general form of $F(z)$ and $G(z)$ is a nonperturbative effect, so one can also expect an increase of the number of physical degrees of freedom, as well as the appearance of pathologies such as tachyons (one can also recall that at least in the scalar field theory higher derivative theories suffer from ghosts; see \cite{Woodard:2006nt} for details). In this connection, it is hard to believe that the perturbation analysis, which is often used to examine fermion sector in brane world models, can adequately describe the theory in the general case, taking into account that it leads to equations of motion which are even more complicated than simple equation (\ref{4Dsceq}) (see also a discussion of this problem in \cite{Smolyakov:2015zsa}). Indeed, the perturbation theory can provide solutions for some of the physical degrees of freedom of the theory, exactly as it happens with the first root in (\ref{4Dperturbroots}), which can be obtained perturbatively. However, the rest of the possible physical degrees of freedom may appear to be lost when one uses perturbation theory in such nontrivial cases.

Of course, there can be some non-obvious ways to solve such fourth-order equation of motion or even to avoid them (for example, by taking some nonstandard combination of the five-dimensional fields such as $p_{1}^{L,R}(z)\Psi_{1}^{L,R}(x,z)+p_{2}^{L,R}(z)\Psi_{2}^{L,R}(x,z)$, where $p_{1}^{L,R}(z)$, $p_{2}^{L,R}(z)$ are some functions), but I have failed to find such possibilities. Thus, the general case with $(e^{\sigma}G)'\not\equiv 0$ and $F(z)\not\equiv 0$, naively leading to equations of form (\ref{hder1}) and (\ref{hder2}), should be carefully and thoroughly examined before considering its phenomenological consequences.

An important remark is in order here. Equations (\ref{hder1}) and (\ref{hder2}) are still fourth-order differential equations even for $F(z)\equiv 0$, whereas it was shown above that the system of equations (\ref{KGfeqs1}) and (\ref{KGfeqs2}) can be decoupled for $F(z)\equiv 0$, leading to the second-order equations of motion for the combinations $\Psi_{1}+\Psi_{2}$ and $\Psi_{1}-\Psi_{2}$ of the fields. This would-be contradiction can be easily resolved. Indeed, for $F(z)\equiv 0$ equations (\ref{hder1}) and (\ref{hder2}) take the form
\begin{equation}\label{hder1a}
\hat L\frac{\hat L}{e^{-\sigma}(e^{\sigma}G)'}\Psi_{1,2}^{L}-e^{-\sigma}(e^{\sigma}G)'\Psi_{1,2}^{L}=0.
\end{equation}
Let us define the operator
\begin{equation}
\hat B=\frac{\hat L}{e^{-\sigma}(e^{\sigma}G)'}.
\end{equation}
Using this definition, equation (\ref{hder1a}) can be rewritten as
\begin{equation}\label{hder1b}
(\hat B-1)(\hat B+1)\Psi_{1,2}^{L}=0.
\end{equation}
The operators $\hat B-1$ and $\hat B+1$ commute, which means that solutions to equation (\ref{hder1b}) are just the linear combinations of solutions to the second-order differential equations $(\hat B-1)\Psi_{a}^{L}=0$ and $(\hat B+1)\Psi_{b}^{L}=0$. The latter equations are nothing but the equations for the combinations $\Psi_{1}+\Psi_{2}$ and $\Psi_{1}-\Psi_{2}$, which appear when one adds and subtracts equations (\ref{KGfeqs1}) and (\ref{KGfeqs2}) with $F(z)\equiv 0$. As we will see in the next section, in fact both equations lead to the same wave functions of the Kaluza-Klein modes.

Of course, a fully analogous procedures can be made for the cases (\ref{partic1}), (\ref{partic2}), but now not in the whole space-time, but only in the regions $0<z<L$ and $-L<z<0$ separately.

An analogous reduction exists in the simple four-dimensional example (\ref{4Dscact}) too. For $\alpha=-\frac{1}{4}$ we get from (\ref{csmasses}) only one root
\begin{equation}
m^{2}=2M^{2},
\end{equation}
equation of motion for the scalar field takes the form
\begin{equation}
(\Box+2M^{2})^2\phi=0.
\end{equation}
This degenerate case does not contain pathologies like tachyons. As will be shown below, the ``degenerate'' case $F\equiv 0$, described by (\ref{hder1b}), also does not provide extra degrees of freedom or pathologies.

\section{Kaluza-Klein decomposition}
Bearing in mind the results, presented in the previous section, we are ready to perform the Kaluza-Klein decomposition. As we will see below, a consistent decomposition indeed demands second-order differential equations for the fields (or for their linear combinations).

Unfortunately, the cases (\ref{partic1}), (\ref{partic2}) appear to be rather complicated and seem not allowing to perform all the calculations analytically in the general case. For example, in the particular case with $\sigma(z)=-k|z|$, $F(z)\equiv Q\,\textrm{sign}(z)$ and $M=0$ in (\ref{partic1}) the mass spectrum appears to be defined by the determinant of a special $4\times 4$ matrix, whereas some coefficients in the wave functions of the modes are defined by eigenvectors of this matrix. The mass spectrum was examined numerically for different choices of the constants $Q$, $\gamma$, providing the normal spectra without tachyons or other pathologies. However, the necessity for numerical calculations makes the whole analysis very complicated. So, below I will consider only those cases, which admit analytical treatment in the general case and are of particular interest. They are:
\begin{enumerate}
\item $G(z)\equiv Me^{-\sigma(z)}$; any $F(z)$.
\item $F(z)\equiv 0$; any $G(z)$.
\item $G(z)\equiv K\,\delta(z-L)$; any $F(z)$.
\end{enumerate}
As we will see below, all these cases demand a considerably different treatment.

\subsection{$G(z)\equiv Me^{-\sigma(z)}$, any $F(z)$}
In metric (\ref{backgmetric}) and with equation (\ref{finetune}), action (\ref{faction}) takes the form\footnote{The fermion action exactly of form (\ref{fermact}) (but in other notations) was considered in \cite{CG} for examining discrete symmetries in brane world models.}
\begin{eqnarray}\nonumber
S=\int d^{4}xdze^{4\sigma}\Bigl(e^{-\sigma}i\bar\Psi_{1}\gamma^{\mu}\partial_{\mu}\Psi_{1}-\bar\Psi_{1}\gamma^{5}\left(\partial_{5}+2\sigma'\right)\Psi_{1}
\\ \nonumber
+e^{-\sigma}i\bar\Psi_{2}\gamma^{\mu}\partial_{\mu}\Psi_{2}-\bar\Psi_{2}\gamma^{5}\left(\partial_{5}+2\sigma'\right)\Psi_{2}\\ \label{fermact}
-F(z)\left(\bar\Psi_{1}\Psi_{1}-\bar\Psi_{2}\Psi_{2}\right)-Me^{-\sigma}\left(\bar\Psi_{2}\Psi_{1}+\bar\Psi_{1}\Psi_{2}\right)\Bigr),
\end{eqnarray}
where $M>0$. The second-order equations of motion for the components of the fields $\Psi_{1}$ and $\Psi_{2}$ take the form (\ref{KGfeqs3}), (\ref{KGfeqs4}).

It is convenient to represent the five-dimensional fields as \cite{Smolyakov:2011hv}
\begin{eqnarray}\label{dec1}
\Psi_{1}&=&\sum\limits_{n=0}^{\infty}\left(f_n(z)\psi_{L}^n(x)+\frac{{\tilde
f_n(z)}}{D_{n}}\hat\psi_{R}^n(x)\right)\\ \label{dec2}
\Psi_{2}&=&\sum\limits_{n=0}^{\infty}\left(f_n(z)\psi_{R}^n(x)-\frac{{\tilde
f_n(z)}}{D_{n}}\hat\psi_{L}^n(x)\right),
\end{eqnarray}
where the constants $D_{n}$ are introduced for convenience and will be defined later. According to the orbifold symmetry conditions (\ref{sym1}), (\ref{sym2}) for the fields $\Psi_{1}$ and $\Psi_{2}$, the functions $f_n(z)$ and $\tilde f_n(z)$ satisfy the symmetry conditions
\begin{equation}\label{symftildef}
f_n(-z)=f_n(z),\qquad \tilde f_n(-z)=-\tilde f_n(z).
\end{equation}
Substituting the decomposition into equations (\ref{KGfeqs3}), (\ref{KGfeqs4}), we get the following equations for the wave functions $f_n(z)$, ${\tilde
f_n(z)}$:
\begin{eqnarray}\label{e2}
e^{-2\sigma}(m_{n}^2-M^2)f_n+f_n''+5\sigma'f_n'+(6{\sigma'}^2+2\sigma'')f_n-(F^{2}-\sigma'F-F')f_n=0,\\ \label{e3}
e^{-2\sigma}(m_{n}^2-M^2){\tilde f_n}+{\tilde f_n}''+5\sigma'{\tilde f_n}'+(6{\sigma'}^2+2\sigma''){\tilde f_n}-(F^{2}+\sigma'F+F'){\tilde f_n}=0.
\end{eqnarray}
Since all the coefficients in (\ref{e2}) and (\ref{e3}) are all even in $z$ and real, we can always find real solutions to these equations satisfying (\ref{symftildef}).

At this stage it is unclear why we use the same eigenvalues $m_n$ in equations (\ref{e2}) and (\ref{e3}) --- the equations are different, so one can expect that in the general case they can provide different sets of eigenvalues. However, it is easy to show that whenever equations (\ref{e2}), (\ref{e3}) hold, the following system of equations also holds (see Appendix~A):
\begin{eqnarray}\label{e4}
&f_n'+(2\sigma'+F)f_n=(m_n+M)e^{-\sigma}{\tilde f_n},\\
\label{e5} &{\tilde f_n}'+(2\sigma'-F){\tilde f_n}=-(m_n-M)e^{-\sigma}f_n,
\end{eqnarray}
where, without loss of generality, we take $m_{n}\ge 0$.\footnote{By changing $M\to -M$ or/and $m_{n}\to-m_{n}$ one can construct other systems of first-order equations, satisfying (\ref{e2}) and (\ref{e3}), but for $n\ne 0$ these systems of equations can be easily brought back to the form (\ref{e4}), (\ref{e5}), thus not providing any additional solutions.} It means that the eigenfunctions of equations (\ref{e2}), (\ref{e3}) are indeed connected and provide the same set of eigenvalues. For example, given a symmetric solution $f_{n}(z)$ to equation (\ref{e2}), using (\ref{e4}) we can always find the corresponding antisymmetric solution ${\tilde f_{n}}(z)$ to equation (\ref{e3}). As for the zero mode, equations (\ref{e4}), (\ref{e5}) indeed provide $m_{0}=M$, $f_{0}(z)=Ce^{-\int_{0}^{z}F(y)dy}$ and ${\tilde f_{0}(z)}=0$, resulting in (\ref{zm1}), (\ref{zm2}). One can check that the solution for the zero mode $f_{0}(z)$ satisfies the initial equation (\ref{e2}) with $m_{0}=M$. In fact, for the zero mode we can simply use the equations
\begin{equation}
f_{0}'+(2\sigma'+F)f_{0}=0,\qquad {\tilde f_{0}}\equiv 0
\end{equation}
with $m_{0}=M$ from the very beginning.

With the help of (\ref{e2}) and (\ref{e3}) it is possible to show that the orthogonality conditions
\begin{equation}
\int\limits_{-L}^{L}e^{3\sigma} f_{n}f_{k} dz=0,\qquad \int\limits_{-L}^{L}e^{3\sigma}{\tilde f_{n}}{\tilde f_{k}}dz=0,
\qquad n\ne k.
\end{equation}
are fulfilled, whereas $\int\limits_{-L}^{L}e^{3\sigma} f_{n}{\tilde f_{k}} dz=0$ for all $n$ and $k$ because of (\ref{symftildef}). It is convenient to impose the normalization condition $\int\limits_{-L}^{L}e^{3\sigma} f_n^2 dz=1$ and to define $D_{n}^{2}=\int\limits_{-L}^{L}e^{3\sigma}{\tilde f_n}^{2}dz$. The latter results in $D_{n}^{2}=\frac{m_n-M}{m_n+M}$ (the value of $D_n$ can be obtained by integrating the product of equation (\ref{e4}) and $\tilde f_n$ by parts and using equation (\ref{e5})).

Substituting decomposition (\ref{dec1}), (\ref{dec2}) into action (\ref{fermact}), taking into account (\ref{e4}), (\ref{e5}) (these equations are necessary to transform the terms with the function $F(z)$ and with the derivative in the coordinate of the extra dimension in (\ref{fermact}) into the ``mass term'' form) and then integrating over the coordinate of the extra dimension, we get the effective four-dimensional action
\begin{eqnarray}\nonumber
S_{eff}=\int d^{4}x \Biggl(i\bar\psi\gamma^{\mu}\partial_{\mu}\psi-M\bar\psi\psi+\sum\limits_{n=1}^{\infty}\left(
i\bar\psi_{n}\gamma^{\mu}\partial_{\mu}\psi_{n}+i\bar{\hat\psi}_{n}\gamma^{\mu}\partial_{\mu}\hat\psi_{n}\right.\\
\label{effactF}\left.-
M({\bar\psi}_{n}\psi_{n}-\bar{\hat\psi}_{n}\hat\psi_{n})-\sqrt{m_{n}^{2}-M^2}\left(\bar\psi_{n}{\hat\psi}_{n}+
\bar{\hat\psi}_{n}\psi_{n}\right)\right)\Biggr),
\end{eqnarray}
where $\psi=\psi_{L}^0+\psi_{R}^0$ and $\psi_n=\psi_{L}^n+\psi_{R}^n$,
$\hat\psi_n=\hat\psi_{L}^n+\hat\psi_{R}^n$ for $n\ge 1$. All the calculations are straightforward, though rather bulky, and we do not present them here. To diagonalize the mass matrix, we will use the rotation \cite{Smolyakov:2011hv,Smolyakov:2012ud}
\begin{eqnarray}\label{rotation1}
\psi_n(x)&=&\psi_{1,n}(x)\cos(\theta_n)+\psi_{2,n}(x)\sin(\theta_n),\\
\label{rotation2}
\hat\psi_n(x)&=&\psi_{1,n}(x)\sin(\theta_n)-\psi_{2,n}(x)\cos(\theta_n)
\end{eqnarray}
for $n\ge 1$, where $\tan(2\theta_n)=\frac{\sqrt{m_{n}^{2}-M^2}}{M}$, which is fully analogous to (\ref{rotation01}), (\ref{rotation02}). We obtain
\begin{eqnarray}\nonumber
&&S_{eff}=\int d^{4}x\biggl(i\bar\psi\gamma^{\mu}\partial_{\mu}\psi-
M\bar{\psi}\psi\\ &&+
\sum\limits_{n=1}^{\infty}\left(i\bar\psi_{1,n}\gamma^{\mu}\partial_{\mu}\psi_{1,n}+i\bar\psi_{2,n}\gamma^{\mu}\partial_{\mu}\psi_{2,n}-
m_{n}\bar{\psi}_{1,n}\psi_{1,n}+m_{n}\bar{\psi}_{2,n}\psi_{2,n}\right)\biggr).
\end{eqnarray}
The last step is to make the redefinition $\psi_{2,n}\to \gamma^{5}\psi_{2,n}$ in
order to get the conventional sign of the mass term of the
four-dimensional fermion $\psi_{2,n}$. Finally, we get
\begin{eqnarray}
S_{eff}=\int d^{4}x\left(i\bar\psi\gamma^{\mu}\partial_{\mu}\psi-
M\bar{\psi}\psi+
\sum\limits_{n=1}^{\infty}\sum\limits_{i=1}^{2}\left(i\bar\psi_{i,n}\gamma^{\mu}\partial_{\mu}\psi_{i,n}-
m_{n}\bar{\psi}_{i,n}\psi_{i,n}\right)\right).
\end{eqnarray}

We see that a consistent Kaluza-Klein decomposition of fermion fields is not so simple as the Kaluza-Klein decomposition for scalar or gauge fields. It is necessary to point out that:
\begin{enumerate}
\item Two steps are important for correctly performing the Kaluza-Klein decomposition in the general case (i.e., for any form of $F(z)$). First, the second-order differential equations (\ref{KGfeqs3}), (\ref{KGfeqs4}) (and, consequently, equations for the eigenvalues and eigenfunctions (\ref{e2}), (\ref{e3})) suggest the appropriate separation of variables (\ref{dec1}), (\ref{dec2}) for the fields $\Psi_{1}$, $\Psi_{2}$ and provide the complete set of possible physical degrees of freedom of the theory. These equations also allow one to check the absence of possible pathologies (exactly in the same way as the Klein-Gordon equation for each component of the ordinary four-dimensional field guarantees that the theory is pathologically-free).

    Second, the system of first-order differential equations (\ref{e4}), (\ref{e5}), which follows from (\ref{e2}), (\ref{e3}), allows one to get rid of the terms with the function $F(z)$ and with the derivative in the coordinate of the extra dimension in (\ref{fermact}) and to obtain the four-dimensional effective action in a consistent way.
\item At each Kaluza-Klein level with $n\ge 1$ we have two modes with the same four-dimensional mass, exactly as in the flat case \cite{Macesanu} presented in Section~2.
\item The wave function of the zero mode has the same form for any value of the mass $M$, including the case $M=0$, see equations (\ref{zm1}) and (\ref{zm2}).
\item For the flat case with $F(z)\equiv 0$ the wave functions $f_{n}$ and $\tilde f_{n}$ can be easily obtained from (\ref{e4}) and (\ref{e5}). In this case (\ref{dec1}) and (\ref{dec2}) fully coincide with (\ref{subst1}) and (\ref{subst2}).
\end{enumerate}

Now let us proceed to the case $F(z)\equiv 0$, $(e^{\sigma}G)'\not\equiv 0$, which appears to be more complicated.

\subsection{$F(z)\equiv 0$, any $G(z)$}
\subsubsection{The decomposition}
In metric (\ref{backgmetric}) and for $F(z)\equiv 0$, action (\ref{faction}) takes the form
\begin{eqnarray}\nonumber
S=\int d^{4}xdze^{4\sigma}\Bigl(e^{-\sigma}i\bar\Psi_{1}\gamma^{\mu}\partial_{\mu}\Psi_{1}-\bar\Psi_{1}\gamma^{5}\left(\partial_{5}+2\sigma'\right)\Psi_{1}
\\ \label{fermactG}
+e^{-\sigma}i\bar\Psi_{2}\gamma^{\mu}\partial_{\mu}\Psi_{2}-\bar\Psi_{2}\gamma^{5}\left(\partial_{5}+2\sigma'\right)\Psi_{2}
-G(z)\left(\bar\Psi_{2}\Psi_{1}+\bar\Psi_{1}\Psi_{2}\right)\Bigr),
\end{eqnarray}
where $G(-z)=G(z)$ according to the orbifold symmetry conditions and we suppose that $G(z)\ge 0$ for any $z$. Recall that the condition $(e^{\sigma}G)'\equiv 0$ is not supposed to fulfill.

From the very beginning it is convenient to use the combinations
\begin{eqnarray}\label{Psia}
\Psi_{A}=\frac{1}{\sqrt{2}}\left(\Psi_{1}+\Psi_{2}\right),\\ \label{Psib}
\Psi_{B}=\frac{1}{\sqrt{2}}\left(\Psi_{1}-\Psi_{2}\right).
\end{eqnarray}
With these notations equations of motion, following from (\ref{KGfeqs1}), (\ref{KGfeqs2}) with $F(z)\equiv 0$, can be rewritten as
\begin{eqnarray}\label{KGfeqs1G}
-e^{-2\sigma}\Box\Psi_{A}+\Psi_{A}''+5\sigma'\Psi_{A}'+(6{\sigma'}^2+2\sigma'')\Psi_{A}-G^{2}\Psi_{A}+(G'+\sigma'G)\gamma^{5}\Psi_{A}=0,\\
\label{KGfeqs2G}
-e^{-2\sigma}\Box\Psi_{B}+\Psi_{B}''+5\sigma'\Psi_{B}'+(6{\sigma'}^2+2\sigma'')\Psi_{B}-G^{2}\Psi_{B}-(G'+\sigma'G)\gamma^{5}\Psi_{B}=0.
\end{eqnarray}
Solutions to these equations, corresponding to a four-dimensional mass $m$, can be represented as
\begin{eqnarray}\label{psia1}
\Psi_{A}=f(z)\psi_{L}^{A}(x)+f(-z)\psi_{R}^{A}(x),\\ \label{psia2}
\Psi_{B}=f(z)\psi_{R}^{B}(x)+f(-z)\psi_{L}^{B}(x).
\end{eqnarray}
Indeed, the equation for the function $f(z)$, corresponding to the four-dimensional mass $m$, takes the form
\begin{equation}\label{secorderG}
e^{-2\sigma}m^2f(z)+f''(z)+5\sigma'f'(z)+(6{\sigma'}^2+2\sigma'')f(z)-(G^{2}-\sigma'G-G')f(z)=0,
\end{equation}
whereas it is clear that, due to the symmetry properties of $G(z)$, the function $f(-z)$ satisfies the equation
\begin{equation}\label{secorderGminus}
e^{-2\sigma}m^2f(-z)+f''(-z)+5\sigma'f'(-z)+(6{\sigma'}^2+2\sigma'')f(-z)-(G^{2}+\sigma'G+G')f(-z)=0.
\end{equation}
According to the general theory \cite{CL}, the solutions to equation (\ref{secorderG}) (and to equation (\ref{secorderGminus}) too) make up an orthonormal set of eigenfunctions $f_{n}(z)$, the lowest eigenvalue $m_{0}$ being simple. Moreover, it is not difficult to show that $m_{n}^2\ge 0$ for (\ref{secorderG}) (see Appendix~B). The sets of eigenvalues of equations (\ref{secorderG}) and (\ref{secorderGminus}) obviously coincide. The coefficients in equations (\ref{secorderG}) and (\ref{secorderGminus}) are real, so we can consider only real solutions for $f(z)$.

Equations (\ref{secorderG}) and (\ref{secorderGminus}) look similar to equations (\ref{e2}) and (\ref{e3}) (up to the change $F(z)\to G(z)$), however, they correspond to different systems: first, the symmetry properties of the functions $F(z)$ and $G(z)$ are different; and second, equations (\ref{e2}) and (\ref{e3}) correspond to the fields $\Psi_{1}$, $\Psi_{2}$ themselves, whereas equation (\ref{secorderG}) and (\ref{secorderGminus}) correspond to the combinations $\Psi_{1}+\Psi_{2}$, $\Psi_{1}-\Psi_{2}$ of the initial fields.

Now let us discuss the auxiliary first-order equations, which will be necessary for obtaining the four-dimensional effective action. One can show that any solution to equation
\begin{eqnarray}\label{1dim01}
f'(z)+2\sigma'(z)f(z)+G(z)f(z)=me^{-\sigma(z)}f(-z)
\end{eqnarray}
or equation
\begin{eqnarray}\label{1dim02}
f'(z)+2\sigma'(z)f(z)+G(z)f(z)=-me^{-\sigma(z)}f(-z)
\end{eqnarray}
satisfy equation (\ref{secorderG}),\footnote{For double eigenvalues the opposite is not correct --- not any solutions to equation (\ref{secorderG}) satisfy equation (\ref{1dim01}) or (\ref{1dim02}). For example, for $\sigma(z)\equiv 0$ and $G(z)\equiv M$ the functions $f(z)\sim\cos\left(\frac{\pi n}{L}z\right)$, $f(z)\sim\sin\left(\frac{\pi n}{L}z\right)$, $n\ge 1$ are solutions to (\ref{secorderG}) with $m_{n}=\sqrt{\frac{\pi^2 n^2}{L^2}+M^2}$, but not solutions to (\ref{1dim01}) or (\ref{1dim02}) with $\sigma(z)\equiv 0$ and $G(z)\equiv M$. The corresponding orthogonal solutions are: $f(z)\sim\cos\left(\frac{\pi n}{L}z\right)+\sqrt{\frac{m_{n}-M}{m_{n}+M}}\sin\left(\frac{\pi n}{L}z\right)$ for (\ref{1dim01}) and $f(z)\sim\cos\left(\frac{\pi n}{L}z\right)-\sqrt{\frac{m_{n}+M}{m_{n}-M}}\sin\left(\frac{\pi n}{L}z\right)$ for (\ref{1dim02}).} the proof is fully analogous to the one presented in Appendix~A. Please pay attention to the argument $-z$ in the r.h.s. terms of equations (\ref{1dim01}) and (\ref{1dim02}). Equation (\ref{1dim02}) can not be brought to the form (\ref{1dim01}) by a simple redefinition of the function $f(z)$, so in principle we should take {\em both equations}. Let us define solutions to equation (\ref{1dim01}) with eigenvalue $m_{n,1}>0$ as $f_{n,1}(z)$, $n\ge 1$; solutions to equation (\ref{1dim02}) with eigenvalue $m_{n,2}>0$ as $f_{n,2}(z)$, $n\ge 1$. For the zero mode we take only equation (\ref{1dim01}) with the eigenvalue $m_{0}$ and the eigenfunction $f_{0}(z)$. Indeed, the lowest eigenvalue is simple, so we expect only one solution. Since for $G(z)\equiv Me^{-\sigma}$ equation (\ref{1dim01}) gives $m_{0}=M$ and $f_{0}(z)\sim e^{-2\sigma}$, this choice is justified.\footnote{Equation $f'_{0}(z)+2\sigma'(z)f_{0}(z)+G(z)f_{0}(z)=-m_{0}e^{-\sigma(z)}f_{0}(-z)$ for the zero mode is not excluded in principle. However, as we will see below, in the cases in which the perturbation theory can be used the condition $G(z)\ge 0$ and equation (\ref{1dim01}) also give $m_{0}>0$, so below we will use equation (\ref{1dim01}) (i.e., equation with $+m_{0}$) for the zero mode.}

Now let us recall that the initial five-dimensional fields are $\Psi_{1}$, $\Psi_{2}$ and they should satisfy the orbifold symmetry conditions (\ref{sym1}) and (\ref{sym2}). Thus, using (\ref{psia1}) and (\ref{psia2}) we arrive at the Kaluza-Klein decomposition
\begin{eqnarray}\label{KKpsi1}
\Psi_{1}=\frac{1}{\sqrt{2}}\left(f_{+,0}(z)\psi_{L}(x)-f_{-,0}(z)\psi_{R}(x)+\sum\limits_{n=1}^{\infty}\sum\limits_{k=1}^{2}\left(f_{+,n,k}(z)\psi_{L}^{n,k}(x)-f_{-,n,k}(z)\psi_{R}^{n,k}(x)\right)\right),\\ \label{KKpsi2}
\Psi_{2}=\frac{1}{\sqrt{2}}\left(f_{+,0}(z)\psi_{R}(x)+f_{-,0}(z)\psi_{L}(x)+\sum\limits_{n=1}^{\infty}\sum\limits_{k=1}^{2}\left(f_{+,n,k}(z)\psi_{R}^{n,k}(x)+f_{-,n,k}(z)\psi_{L}^{n,k}(x)\right)\right),
\end{eqnarray}
where $f_{+,0}(z)=f_{0}(z)+f_{0}(-z)$, $f_{-,0}(z)=f_{0}(z)-f_{0}(-z)$, $f_{+,n,k}(z)=f_{n,k}(z)+f_{n,k}(-z)$, $f_{-,n,k}(z)=f_{n,k}(z)-f_{n,k}(-z)$. The Kaluza-Klein decomposition for the combinations $\Psi_{A}$, $\Psi_{B}$ now looks like
\begin{eqnarray}\label{KKpsia}
\Psi_{A}=f_{0}(z)\psi_{L}(x)+f_{0}(-z)\psi_{R}(x)+\sum\limits_{n=1}^{\infty}\sum\limits_{k=1}^{2}\left(f_{n,k}(z)\psi_{L}^{n,k}(x)+f_{n,k}(-z)\psi_{R}^{n,k}(x)\right),\\ \label{KKpsib} \Psi_{B}=-f_{0}(z)\psi_{R}(x)+f_{0}(-z)\psi_{L}(x)+\sum\limits_{n=1}^{\infty}\sum\limits_{k=1}^{2}\left(-f_{n,k}(z)\psi_{R}^{n,k}(x)+f_{n,k}(-z)\psi_{L}^{n,k}(x)\right),
\end{eqnarray}

The corresponding first-order equations, which will be necessary for obtaining the effective four-dimensional action, take the form:
\begin{equation}\label{1dim0}
f'_{0}(z)+2\sigma'(z)f_{0}(z)+G(z)f_{0}(z)=m_{0}e^{-\sigma(z)}f_{0}(-z)
\end{equation}
for the zero mode and
\begin{eqnarray}\label{1dim1}
&&f'_{n,1}(z)+2\sigma'(z)f_{n,1}(z)+G(z)f_{n,1}(z)=m_{n,1}e^{-\sigma(z)}f_{n,1}(-z),\\ \label{1dim2}
&&f_{n,2}'(z)+2\sigma'(z)f_{n,2}(z)+G(z)f_{n,2}(z)=-m_{n,2}e^{-\sigma(z)}f_{n,2}(-z)
\end{eqnarray}
for the modes with $n\ge 1$. Here $m_{0}\ge 0$, $m_{n,k}>0$. Since solutions to equation (\ref{1dim0}), (\ref{1dim1}) and (\ref{1dim2}) satisfy equation (\ref{secorderG}), it is clear that for the modes, corresponding to different eigenvalues, the following orthogonality conditions hold:
\begin{eqnarray}\label{orth1}
\int\limits_{-L}^{L}e^{3\sigma}f_{n,k}(z)f_{j,l}(z)dz=\int\limits_{-L}^{L}e^{3\sigma}f_{n,k}(-z)f_{j,l}(-z)dz=0,&&\quad m_{n,k}\ne m_{j,l},\\ \label{orth4}
\int\limits_{-L}^{L}e^{3\sigma}f_{0}(z)f_{n,k}(z)dz=\int\limits_{-L}^{L}e^{3\sigma}f_{0}(-z)f_{n,k}(-z)dz=0,&&\quad n\ge 1.
\end{eqnarray}

Now we are ready to obtain the four-dimensional effective action for the physical degrees of freedom of the theory. First, it is convenient to take the normalization conditions
\begin{equation}\label{norm1and2ext}
\int\limits_{-L}^{L}e^{3\sigma}f_{0}^{2}(z)dz=\int\limits_{-L}^{L}e^{3\sigma}f_{n,1}^{2}(z)dz=\int\limits_{-L}^{L}e^{3\sigma}f_{n,2}^{2}(z)dz=\frac{1}{2}.
\end{equation}
Another useful step is to express action (\ref{fermactG}) through the combinations $\Psi_{A}$ (\ref{Psia}) and $\Psi_{B}$ (\ref{Psib}) as
\begin{eqnarray}\nonumber
S=\int d^{4}xdze^{4\sigma}\Bigl(e^{-\sigma}i\bar\Psi_{A}\gamma^{\mu}\partial_{\mu}\Psi_{A}-\bar\Psi_{A}\gamma^{5}\left(\partial_{5}+2\sigma'\right)\Psi_{A}
\\ \label{fermactGpsiab}
+e^{-\sigma}i\bar\Psi_{B}\gamma^{\mu}\partial_{\mu}\Psi_{B}-\bar\Psi_{B}\gamma^{5}\left(\partial_{5}+2\sigma'\right)\Psi_{B}
-G(z)\left(\bar\Psi_{A}\Psi_{A}-\bar\Psi_{B}\Psi_{B}\right)\Bigr).
\end{eqnarray}
Substituting the Kaluza-Klein decomposition (\ref{KKpsia}) and (\ref{KKpsib}) into (\ref{fermactGpsiab}), using equations (\ref{1dim0}), (\ref{1dim1}) and (\ref{1dim2}) to transform the terms with the function $G(z)$ and with the derivative in the coordinate of the extra dimension in (\ref{fermactGpsiab}) into the terms containing $m_{0}$ or $m_{n,k}$, using the orthogonality conditions (\ref{orth1})--(\ref{orth4}) and the normalization conditions (\ref{norm1and2ext}) when integrating over the coordinate of the extra dimension $z$, we arrive at
\begin{eqnarray}\nonumber
S_{eff}=\int d^{4}x\Biggl(i\bar\psi\gamma^{\mu}\partial_{\mu}\psi-m_{0}\bar\psi\psi\\ \label{effactG}+\sum\limits_{n=1}^{\infty}\Bigl(
i\bar\psi_{n,1}\gamma^{\mu}\partial_{\mu}\psi_{n,1}+i\bar\psi_{n,2}\gamma^{\mu}\partial_{\mu}\psi_{n,2}-m_{n,1}\bar\psi_{n,1}\psi_{n,1}+m_{n,2}\bar\psi_{n,2}
\psi_{n,2}\Bigr)\Biggr),
\end{eqnarray}
where $\psi=\psi_{L}+\psi_{R}$, $\psi_{n,1}=\psi_{L}^{n,1}+\psi_{R}^{n,1}$, $\psi_{n,2}=\psi_{L}^{n,2}+\psi_{R}^{n,2}$. The final step is to make the redefinition $\psi_{n,2}\to \gamma^{5}\psi_{n,2}$ in order to get the conventional sign of the mass terms of the four-dimensional fermions $\psi_{n,2}$. Finally, we get
\begin{equation}\label{effactGabfinal}
S_{eff}=\int d^{4}x\Biggl(i\bar\psi\gamma^{\mu}\partial_{\mu}\psi-m_{0}\bar\psi\psi+\sum\limits_{n=1}^{\infty}\sum\limits_{k=1}^{2}\left(
i\bar\psi_{n,k}\gamma^{\mu}\partial_{\mu}\psi_{n,k}-m_{n,k}\bar\psi_{n,k}\psi_{n,k}\right)\Biggr).
\end{equation}

It is necessary to point out that:
\begin{enumerate}
\item As in the previous case, the second-order differential equations (\ref{KGfeqs1G}), (\ref{KGfeqs2G}) (and (\ref{secorderG})) suggest the appropriate separation of variables (\ref{KKpsi1}), (\ref{KKpsi2}) for the fields $\Psi_{1}$, $\Psi_{2}$, provide the complete set of the possible physical degrees of freedom of the theory and allow one to check the absence of possible pathologies; whereas first-order differential equations (\ref{1dim0})--(\ref{1dim2}) allow one to get rid of the terms with the function $G(z)$ and with the derivative in the coordinate of the extra dimension in (\ref{fermact}) and to obtain the four-dimensional effective action in a consistent way.
\item The wave function of the zero mode and the value of the four-dimensional mass of this mode now depend on the form of function $G(z)$.
\end{enumerate}

An important remark is in order here. One can expect that in the most cases all the eigenvalues $m_{n,k}$, $n\ge 1$ are simple. Indeed, it was noted in \cite{DiffEq} that in general the double eigenvalues are not common, whereas the Fourier case is not typical. However, the double eigenvalues are still possible, the simplest example with $\sigma(z)\equiv 0$, $G(z)\equiv M$ (the Fourier case) has already been discussed in Section~2. Moreover, action (\ref{fermactG}) with $G(z)\equiv Me^{-\sigma}$ and action (\ref{fermact}) with $F(z)\equiv 0$ correspond to the same five-dimensional Lagrangian, whereas the results of Subsection~4.1 suggest that in this case there are two different four-dimensional modes of the same mass at each Kaluza-Klein level with $n\ge 1$. So, let us discuss this point in more detail.

Indeed, equations (\ref{e4}), (\ref{e5}) and (\ref{1dim0})--(\ref{1dim2}) for the wave functions have a completely different form, though in fact both systems of equations correspond to the same five-dimensional action. The difference in the diagonalization of the four-dimensional effective actions (\ref{effactF}) and (\ref{effactG}) is also obvious: in the first case there exists a nonzero rotation angle $\theta_{n}$, depending on the number of Kaluza-Klein level $n$, whereas no rotation is necessary in the second case. However, in both cases for the zero mode we get $m_{0}=M$, $f_{0}(z)\sim e^{-2\sigma}$. This indication implies that in the case $F(z)\equiv 0$, $G(z)\equiv Me^{-\sigma}$ the Kaluza-Klein decomposition can be performed in two different ways, both of them have already been presented in Section~4. Of course, both ways lead to the same four-dimensional effective theory: for $G(z)\equiv Me^{-\sigma}$ and for $n\ge 1$ solutions to equations (\ref{1dim1}), (\ref{1dim2}) can be expressed trough solutions to equations (\ref{e4}), (\ref{e5}) as
\begin{eqnarray}\label{ftof1}
f_{n,1}(z)\sim f_{n}(z)-\tilde f_{n}(z),\\ \label{ftof2}
f_{n,2}(z)\sim f_{n}(z)+\frac{\tilde f_{n}(z)}{D_{n}^{2}},
\end{eqnarray}
where $D_{n}=\sqrt{\frac{m_n-M}{m_n+M}}$ (see Subsection~4.1). One can check that (\ref{ftof1}) satisfies (\ref{1dim1}), whereas (\ref{ftof2}) satisfies (\ref{1dim2}). So, in the case $G(z)\equiv Me^{-\sigma}$ equations (\ref{1dim1}), (\ref{1dim2}) indeed provide two different solutions with $m_{n,1}=m_{n,2}$. One can check that the orthogonality condition
\begin{equation}
\int\limits_{-L}^{L}e^{3\sigma}f_{n,1}(z)f_{n,2}(z)dz=0
\end{equation}
holds for (\ref{ftof1}), (\ref{ftof2}), so one can perform the Kaluza-Klein decomposition exactly in the same way as for the case with simple eigenvalues $m_{n,k}$, leading to (\ref{effactGabfinal}) with $m_{n,1}=m_{n,2}$. For $G(z)\not\equiv 0$ which is slightly different from $Me^{-\sigma}$ (even if $M=0$) the degeneracy appears to be removed, leading to modes with different, but close values of the four-dimensional mass.

There also arises the question, whether it is possible to find for the case $(e^{\sigma}G)'\not\equiv 0$, $F(z)\equiv 0$ a system of the first-order equations, playing the role of (\ref{1dim0}), (\ref{1dim1}) and (\ref{1dim2}), such that for $G(z)\equiv Me^{-\sigma}$ it would lead directly to the system of equations (\ref{e4}), (\ref{e5}) with $F(z)\equiv 0$. I have failed to find such a system of equations. A possible explanation of this fact is the following: since it seems that there is no such system of first-order differential equations for the general case (recall fourth-order differential equations (\ref{hder1}), (\ref{hder2})), it looks as if the cases $F(z)\not\equiv 0$, $G(z)\equiv Me^{-\sigma}$ and $F(z)\equiv 0$, $(e^{\sigma}G)'\not\equiv 0$ correspond to completely different branches of the theory, demanding a different treatment. So, even at the ``intersection point'', corresponding to $F(z)\equiv 0$, $G(z)\equiv Me^{-\sigma}$, the first-order systems of differential equations formally do not coincide, though can be connected.

\subsubsection{Perturbation theory}
Equations (\ref{1dim0}), (\ref{1dim1}) and (\ref{1dim2}) allows one to examine consistently the case in which the term with $G(z)$ in (\ref{fermactG}) can be considered as a perturbation. Here we will do it for the case of the zero mode only, because there exists an exact analytical solution for the zero mode wave function for $G(z)\equiv 0$ regardless of the explicit form of $\sigma(z)$.

Let us represent $f_{0}(z)$ as
\begin{equation}\label{substpert}
f_{0}(z)=C_{0}e^{-2\sigma}\left(1+g_{+}(z)+g_{-}(z)\right),
\end{equation}
where $C_{0}$ is a normalization constant, $g_{+}(-z)=g_{+}(z)$, $g_{-}(-z)=-g_{-}(z)$. It is clear that $g_{+}(z)\equiv 0$, $g_{-}(z)\equiv 0$, $m_{0}=0$ for $G(z)\equiv 0$. Substituting (\ref{substpert}) into (\ref{1dim0}), combining the terms possessing the same symmetry ($-z\leftrightarrow z$) properties and of the same orders in perturbations, we get
\begin{eqnarray}\label{pertg-}
g_{-}'(z)=m_{0}e^{-\sigma}-G(z),\\ \label{pertg+}
g_{+}'(z)=-\left(G(z)+m_{0}e^{-\sigma}\right)g_{-}(z).
\end{eqnarray}
From these equations it follows that in the leading order in the perturbation
\begin{equation}\label{pertmass}
m_{0}=\left(\int\limits_{-L}^{L}e^{-\sigma}dz\right)^{-1}\int\limits_{-L}^{L}G(z)dz=\left(\int\limits_{0}^{L}e^{-\sigma}dz\right)^{-1}\int\limits_{0}^{L}G(z)dz.
\end{equation}
Of course, formulas (\ref{pertg-}), (\ref{pertg+}) and (\ref{pertmass}) can be obtained directly from (\ref{secorderG}), though in this case their derivation appears to be more bulky. The condition $$|g_{+}(z)+g_{-}(z)|\ll 1$$ should be fulfilled for any $z$.

If $\alpha_{G}\ll 1$ is a small parameter characterising the function $G(z)$ (more precisely, characterising the ratio of the energy scale corresponding to $G(z)$ and the five-dimensional energy scale $\sim\frac{1}{L}$), then $m_{0}\sim\alpha_{G}$, $g_{-}(z)\sim\alpha_{G}$, $g_{+}(z)\sim\alpha_{G}^2$. An explicit solution to equations (\ref{pertg-}), (\ref{pertg+}) such that $g_{+}(-z)=g_{+}(z)$, $g_{-}(-z)=-g_{-}(z)$ looks like
\begin{eqnarray}\label{pertg-sol}
&&g_{-}(z)=\int\limits_{0}^{z}\left(m_{0}e^{-\sigma(y)}-G(y)\right)dy,\\ \label{pertg+sol}
&&g_{+}(z)=-\int\limits_{0}^{z}\left(G(y)+m_{0}e^{-\sigma(y)}\right)\left(\int\limits_{0}^{y}\left(m_{0}e^{-\sigma(t)}-G(t)\right)dt\right)dy+J,
\end{eqnarray}
where $m_{0}$ is defined by (\ref{pertmass}). It is convenient to choose the constant of integration $J$ such that the relation
\begin{equation}\label{Jdef}
\int\limits_{-L}^{L}\left(2g_{+}(z)+g_{-}^{2}(z)\right)dz=0
\end{equation}
holds. In this case the normalization constant $C_{0}=\left(2\int\limits_{-L}^{L}e^{-\sigma}dz\right)^{-\frac{1}{2}}$ (up to and including the corrections of the order of $\alpha_{G}^{2}$). As expected, in the case $G(z)\equiv Me^{-\sigma}$ equations (\ref{pertmass})--(\ref{pertg+sol}) with (\ref{Jdef}) give $g_{+}(z)\equiv 0$, $g_{-}(z)\equiv 0$.

An analogous procedure can be made for other Kaluza-Klein modes, but for these modes one needs the explicit form of the metric to perform the calculations (i.e., to get the unperturbed solutions to equations (\ref{1dim1}), (\ref{1dim2}) with $G(z)\equiv 0$).

An important comment is in order here. Usually the perturbation analysis is performed using the unperturbed wave functions of fermions, i.e., the mass term is taken into account by substituting the unperturbed wave functions into the term
\begin{equation}\label{pertinc}
\int d^{4}xdze^{4\sigma}G(z)\left(\bar\Psi_{2}\Psi_{1}+\bar\Psi_{1}\Psi_{2}\right)
\end{equation}
of the five-dimensional action and then integrating over $z$. In this case for the mass of the zero mode we obtain
\begin{equation}\label{pertincmass}
m_{0}=\int\limits_{-L}^{L} dze^{4\sigma}G(z)2\left(C_{0}e^{-2\sigma}\right)^2=\left(\int\limits_{-L}^{L}e^{-\sigma}dz\right)^{-1}\int\limits_{-L}^{L}G(z)dz.
\end{equation}
This result coincides with (\ref{pertmass}). However, the chiral structure of the zero mode is not taken into account in such an analysis. Indeed, for the five-dimensional fields (with only the zero mode retained) we get in this case from (\ref{KKpsi1}) and (\ref{KKpsi2})
\begin{eqnarray}
\Psi_{1}=\sqrt{2}C_{0}e^{-2\sigma}\psi_{L}(x),\\ \label{pertzminc}
\Psi_{2}=\sqrt{2}C_{0}e^{-2\sigma}\psi_{R}(x),
\end{eqnarray}
whereas the use of the consistent perturbation analysis reveals
\begin{eqnarray}\label{zerom1}
\Psi_{1}=\sqrt{2}C_{0}e^{-2\sigma}\Bigl((1+g_{+}(z))\psi_{L}(x)-g_{-}(z)\psi_{R}(x)\Bigr),\\ \label{zerom2}
\Psi_{2}=\sqrt{2}C_{0}e^{-2\sigma}\Bigl((1+g_{+}(z))\psi_{R}(x)+g_{-}(z)\psi_{L}(x)\Bigr).
\end{eqnarray}
In principle, the contribution of the extra terms in (\ref{zerom1}), (\ref{zerom2}) is of the order of $\alpha_{G}\sim m_{0}L\ll 1$, whereas the corrections in the effective theory caused by these extra terms are expected to be of the order of $\alpha_{G}^{2}$ and can be neglected in many cases. However, an analogous modification of the gauge boson wave functions leads to severe restrictions on the size of the extra dimension in the Randall-Sundrum model \cite{Csaki:2002gy,Burdman:2002gr}, so one can think that the modification of the fermion wave functions may also lead to analogous effects (this point was also briefly discussed in \cite{Smolyakov:2015zsa}), especially in the case of heavy fermions.

\subsubsection{Higgs field on the brane}
Here we briefly discuss the scenario, in which the Higgs field is supposed to be located exactly on the brane (a detailed analysis of this scenario will be presented in the next subsection). This case can be reproduced by considering the function $G(z)$ to be, for example, $G(z)\sim\delta(z-L)$. Of course, if the delta-function is supposed to be only an approximation of some function, which has a very narrow but nonzero profile in the vicinity of $z=L$, one can simply apply the results presented above to this case. Here we consider the case, in which $G(z)$ is proportional to the {\em exact} delta-function in the mathematical sense.

Let us take equations (\ref{feqs1}), (\ref{feqs2}) with $F(z)\equiv 0$, $G(z)\equiv K\delta(z-L)$, where $K>0$ is a dimensionless constant (in a particular model this constant should be identified with the vacuum expectation value of the Higgs field on the brane made dimensionless by an appropriate five-dimensional energy scale parameter). According to the form of these equations, the terms with delta-functions can be compensated only by the terms with derivative $\partial_{5}$. But this can happen only if the fields $\Psi_{1}$ and $\Psi_{2}$ are {\em discontinuous} at $z=L$. Although it looks rather unnatural, we will show that this discontinuity is not a problem from the mathematical point of view.

To examine this case, let us consider the flat background metric $\sigma(z)\equiv 0$ (just for simplicity: it allows one to obtain the wave function explicitly) and retain only the zero Kaluza-Klein mode. Then the solution to equation
(\ref{1dim0}) with $\sigma(z)\equiv 0$ and $G(z)\equiv K\delta(z-L)$ is (up to a normalization constant)
\begin{equation}\label{discont}
f_{0}(z)=\cos(m_{0}z)+\sin(m_{0}z),\qquad \tan(m_{0}L)=\frac{K}{2}.
\end{equation}
In the derivation of this solution, the term $\sin(m_{0}z)\delta(z-L)$ was handled as
\begin{eqnarray}\nonumber
&&\sin(m_{0}z)\,\delta(z-L)=-\sin(m_{0}L)\,\textrm{sign}(z-L)\,\delta(z-L)\\ \label{regsigndelta}&&=-\frac{1}{2}\sin(m_{0}L)\,\textrm{sign}(z-L)\,\frac{d}{dz}\textrm{sign}(z-L)
=-\frac{1}{4}\sin(m_{0}L)\,\frac{d}{dz}\textrm{sign}^{2}(z-L)=0,
\end{eqnarray}
where we have used the regularization $\textrm{sign}^{2}(z-L)=1$. We see that $f_{0}(z)$ is indeed discontinuous at $z=L$.

Now, from (\ref{KKpsi1}) and (\ref{KKpsi2}) we get for the zero mode (again up to a normalization constant)
\begin{eqnarray}
&&\Psi_{1}^{zero}=\cos(m_{0}z)\psi_{L}(x)-\sin(m_{0}z)\psi_{R}(x),\\
&&\Psi_{2}^{zero}=\cos(m_{0}z)\psi_{R}(x)+\sin(m_{0}z)\psi_{L}(x).
\end{eqnarray}
Substituting this representation into the equations of motion (\ref{feqs1}), (\ref{feqs2}) with $\sigma(z)\equiv 0$, $F(z)\equiv 0$, $G(z)\equiv K\delta(z-L)$, we obtain
\begin{equation}
i\gamma^{\mu}\partial_{\mu}\psi_{L}-m_{0}\psi_{R}=0,\qquad i\gamma^{\mu}\partial_{\mu}\psi_{R}-m_{0}\psi_{L}=0,
\end{equation}
where $m_{0}$ is defined by (\ref{discont}), describing the standard four-dimensional Dirac fermion of mass $m_{0}$. Note that although equation (\ref{1dim0}) is connected with equation (\ref{secorderG}), which does not imply the use of delta-functions, equation (\ref{1dim0}) with $G(z)\equiv K\delta(z-L)$ provides a correct solution to equations of motion (\ref{feqs1}), (\ref{feqs2}). This happens because one can use any smooth approximation for the delta-function, for example, the Gaussian profile, without breaking the structure of equations of motion. This fact is the main motivation to briefly discuss the case $G(z)\equiv K\delta(z-L)$ before its detailed analysis in the next subsection.

For $K\ll 1$, in the leading order in $K$ we get
\begin{equation}\label{discontpert}
f_{0}(z)\approx 1+\frac{K}{2}\,\frac{z}{L},\qquad m_{0}\approx \frac{K}{2L}.
\end{equation}
We see that the correction to the unperturbed wave function appears to be $\left|\frac{K}{2}\frac{z}{L}\right|\le\frac{K}{2}\ll 1$. Note that formulas (\ref{pertmass}) and (\ref{pertincmass}) also give $m_{0}\approx \frac{K}{2L}$.

As for the ``normal'' case, in which the mass generating term has a very narrow, but nonzero profile, in such a case the function $f_{0}(z)$ would vary very rapidly in the vicinity of $z=L$, but of course it would be continuous at $z=L$. Thus, the discontinuity of $f_{0}(z)$ in the present case can be considered simply as an approximation of such rapidly varying but continuous functions.

\subsection{$G(z)\equiv K\,\delta(z-L)$, any $F(z)$}
Now we consider the third case with the action of the form
\begin{eqnarray}\nonumber
S=\int d^{4}xdze^{4\sigma}\Bigl(e^{-\sigma}i\bar\Psi_{1}\gamma^{\mu}\partial_{\mu}\Psi_{1}-\bar\Psi_{1}\gamma^{5}\left(\partial_{5}+2\sigma'\right)\Psi_{1}
+e^{-\sigma}i\bar\Psi_{2}\gamma^{\mu}\partial_{\mu}\Psi_{2}\\ \label{fermactdeltaK}-\bar\Psi_{2}\gamma^{5}\left(\partial_{5}+2\sigma'\right)\Psi_{2}
-F(z)\left(\bar\Psi_{1}\Psi_{1}-\bar\Psi_{2}\Psi_{2}\right)-K\,\delta(z-L)\left(\bar\Psi_{2}\Psi_{1}+\bar\Psi_{1}\Psi_{2}\right)\Bigr),
\end{eqnarray}
where $K>0$ is a constant. It describes the Higgs field located exactly on the brane, but contrary to the case discussed at the end of the previous subsection, here $F(z)\not\equiv 0$. Without loss of generality we take this brane to be the one at $z=L$.

It is clear that everywhere except the point $z=L$ the following five-dimensional equations hold:
\begin{eqnarray}\label{KGfeqs3delta}
-e^{-2\sigma}\Box\Psi_{1}+\Psi_{1}''+5\sigma'\Psi_{1}'+(6{\sigma'}^2+2\sigma'')\Psi_{1}-(F^{2}-e^{-\sigma}(e^{\sigma}F)'\gamma^{5})\Psi_{1}=0,\\
\label{KGfeqs4delta}
-e^{-2\sigma}\Box\Psi_{2}+\Psi_{2}''+5\sigma'\Psi_{2}'+(6{\sigma'}^2+2\sigma'')\Psi_{2}-(F^{2}+e^{-\sigma}(e^{\sigma}F)'\gamma^{5})\Psi_{2}=0.
\end{eqnarray}
These equations are not coupled and suggest the decomposition (for simplicity, we keep only a single mode)
\begin{eqnarray}\label{dec1delta}
\Psi_{1}&=&f\psi_{L}(x)+{\tilde f(z)}\psi_{R}(x)\\ \label{dec2delta}
\Psi_{2}&=&f\hat\psi_{R}(x)-{\tilde f(z)}\hat\psi_{L}(x).
\end{eqnarray}
According to the symmetry conditions, the functions $f(-z)=f(z)$ and $\tilde f(-z)=-\tilde f(z)$ are supposed to satisfy the equations
\begin{eqnarray}\label{e2delta}
&e^{-2\sigma}m^2f+f''+5\sigma'f'+(6{\sigma'}^2+2\sigma'')f-(F^{2}-\sigma'F-F')f=0,\\ \label{e3delta}
&e^{-2\sigma}m^2{\tilde f}+{\tilde f}''+5\sigma'{\tilde f}'+(6{\sigma'}^2+2\sigma''){\tilde f}-(F^{2}+\sigma'F+F'){\tilde f}=0,
\end{eqnarray}
where $\Box\psi_{L,R}+m^{2}\psi_{L,R}=0$ and $\Box\hat\psi_{L,R}+m^{2}\hat\psi_{L,R}=0$. Note that here $m$ is not an eigenvalue, but just a parameter, and the functions $f(z)$ and $\tilde f(z)$ are not eigenfunctions --- equations (\ref{e2delta}) and (\ref{e3delta}) do not hold at $z=L$. So $f(z)$ and $\tilde f(z)$ are just solutions to equations (\ref{e2delta}) and (\ref{e3delta}) with some parameter $m$ which is not defined yet.

Analogously to the previous cases, it is possible to show that whenever equations (\ref{e2delta}), (\ref{e3delta}) hold, the following system of equations also holds:
\begin{eqnarray}\label{e4delta}
&f'(z)+(2\sigma'+F)f(z)=me^{-\sigma}{\tilde f}(z),\\
\label{e5delta} &{\tilde f}'(z)+(2\sigma'-F){\tilde f}(z)=-me^{-\sigma}f(z),
\end{eqnarray}
again everywhere except the point $z=L$. Substituting (\ref{dec1delta}), (\ref{dec2delta}) into equations (\ref{feqs1}), (\ref{feqs2}) with $G(z)\equiv K\delta(z-L)$ and using (\ref{e4delta}), (\ref{e5delta}), we get everywhere except $z=L$
\begin{eqnarray}\label{4Ddiracdelta1}
i\gamma^{\mu}\partial_{\mu}\psi_{L}-m\psi_{R}=0,\qquad
i\gamma^{\mu}\partial_{\mu}\psi_{R}-m\psi_{L}=0,\\ \label{4Ddiracdelta2}
i\gamma^{\mu}\partial_{\mu}\hat\psi_{L}-m\hat\psi_{R}=0,\qquad
i\gamma^{\mu}\partial_{\mu}\hat\psi_{R}-m\hat\psi_{L}=0,
\end{eqnarray}
which are the standard four-dimensional Dirac equations. At the point $z=L$ equations (\ref{feqs1}) and (\ref{feqs2}) give
\begin{eqnarray}\label{bound1delta}
\hat\psi_{R}(x)=-\lim\limits_{\epsilon\to +0}\frac{2{\tilde f(L-\epsilon)}}{Kf(L)}\psi_{R}(x),\\ \label{bound2delta}
\hat\psi_{L}(x)=-\lim\limits_{\epsilon\to +0}\frac{Kf(L)}{2{\tilde f(L-\epsilon)}}\psi_{L}(x)
\end{eqnarray}
for any $x$. In derivation of (\ref{bound1delta}) and (\ref{bound2delta}), we have used the regularization ${\tilde f}(z)\delta(z-L)\sim\textrm{sign}(z-L)\delta(z-L)=0$ (see (\ref{regsigndelta})). Note that in order to support the existence of the delta-function in (\ref{fermactdeltaK}), the function $\tilde f(z)$ should be discontinuous at $z=L$. Finally, substituting (\ref{bound1delta}) and (\ref{bound2delta}) into (\ref{4Ddiracdelta2}), we can find that (\ref{4Ddiracdelta1}) and (\ref{4Ddiracdelta2}) are consistent if the condition
\begin{equation}\label{bounddelta}
\lim\limits_{\epsilon\to +0}\frac{K^{2}f^{2}(L)}{4{\tilde f^{2}(L-\epsilon)}}=1
\end{equation}
holds. This condition defines the allowed values of $m$, i.e., the mass spectrum of the theory.

From the calculations presented above it is not clear what are the possible values of $m^2$, i.e., can $m^2$ be negative or not. Contrary to the previous cases, where the nonnegativity of $m_{n}^2$ follows from the structure of the corresponding second-order differential equations, here we do not have such an equation in the whole extra dimension. However, it can be shown explicitly that the values $m^2<0$ are impossible in the model at hand, see Appendix~C.

Finally, let us derive the effective four-dimensional action of this theory. From (\ref{bounddelta}) it follows that the mass spectrum is defined by
\begin{equation}\label{bounddeltaextra}
\lim\limits_{\epsilon\to +0}\frac{Kf(L)}{2{\tilde f(L-\epsilon)}}=-\beta_{i},\qquad i=1,2
\end{equation}
where
\begin{equation}
\beta_{1}=1,\qquad \beta_{2}=-1.
\end{equation}
Numerical calculations for the case $\sigma(z)\equiv 0$, $F(z)\sim\textrm{sign}(z)$ confirm that there may be solutions to equations (\ref{bounddeltaextra}) in both cases (these numerical calculations are not presented here in order not to overload the text, but they are simple and can be easily reproduced).

Let us label the solutions to equation (\ref{bounddeltaextra}) as $m_{n,i}$ and suppose that all the masses are different (i.e., we do not have degenerate modes, this assumption is also supported by the numerical calculations for the case $\sigma(z)\equiv 0$, $F(z)\sim\textrm{sign}(z)$). The complete decomposition of the five-dimensional fields has the form
\begin{eqnarray}\label{dec1deltafull}
\Psi_{1}&=&f_{0}(z)\psi_{L}(x)+{\tilde
f_{0}(z)}\psi_{R}(x)+\sum\limits_{n=1}^{\infty}\sum\limits_{i=1}^{2}\left(f_{n,i}(z)\psi_{L}^{n,i}(x)+{\tilde
f_{n,i}(z)}\psi_{R}^{n,i}(x)\right),\\ \label{dec2deltafull}
\Psi_{2}&=&\beta_{0}\left(f_{0}(z)\psi_{R}(x)-{\tilde
f_{0}(z)}\psi_{L}(x)\right)+\sum\limits_{n=1}^{\infty}\sum\limits_{i=1}^{2}\beta_{i}\left(f_{n,i}(z)\psi_{R}^{n,i}(x)-{\tilde
f_{n,i}(z)}\psi_{L}^{n,i}(x)\right),
\end{eqnarray}
where we have used (\ref{bound1delta}), (\ref{bound2delta}) and (\ref{bounddeltaextra}), the lowest zero mode is isolated for convenience. Here the modes with $i=1$ are defined by (\ref{bounddeltaextra}) with $\beta_{1}=1$, whereas the modes with $i=2$ are defined by (\ref{bounddeltaextra}) with $\beta_{2}=-1$. As for the zero mode, we do not specify the value of $\beta_{0}$ here --- in principle, it can be equal either to $+1$ or to $-1$.

With (\ref{bounddeltaextra}), it is possible to show that the following orthogonality conditions hold:
\begin{eqnarray}\label{orthdelta}
&&\int\limits_{-L}^{L}e^{3\sigma}\left(f_{n,i}(z)f_{k,i}(z)+{\tilde f}_{n,i}(z){\tilde f}_{k,i}(z)\right)dz=0,\qquad n\neq k,\quad i=1,2,\\
\label{orthdeltamix}
&&\int\limits_{-L}^{L}e^{3\sigma}\left(f_{n,1}(z)f_{k,2}(z)-{\tilde f}_{n,1}(z){\tilde f}_{k,2}(z)\right)dz=0,
\end{eqnarray}
see the detailed derivation in Appendix~D. It is clear that the zero mode can be easily incorporated into these orthogonality conditions (depending on the value of $\beta_{0}$, it should be added to the solutions with $i=1$ or $i=2$). We see that contrary to the case discussed in Subsection~4.1, where the functions, analogous to $f_{n,i}(z)$, $f_{k,j}(z)$ and ${\tilde f}_{n,i}(z)$, ${\tilde f}_{k,j}(z)$, were orthogonal independently, here only the combinations of all four functions give the orthogonality conditions. This happens because here we do not have second-order differential equations valid for all $z$, which usually provide the orthogonality conditions of the standard form.

The last useful step is to find the equation instead of equation (\ref{e5delta}), which is valid in the whole extra dimension. Indeed, equation for the mass spectrum (\ref{bounddeltaextra}) relates the values of $f_{n}(L)$ and ${\tilde f}_{n}(L)$, so we can use it to supply (\ref{e5delta}) by an extra term in order to obtain the systems of equations
\begin{eqnarray}\label{e4deltafullzero}
&&f_{0}'+(2\sigma'+F)f_{0}=m_{0}e^{-\sigma}{\tilde f}_{0},\\ \label{e5deltafullzero}
&&{\tilde f}_{0}'+(2\sigma'-F){\tilde f}_{0}-\frac{K}{\beta_{0}}f_{0}\,\delta(z-L)=-m_{0}e^{-\sigma}f_{0}
\end{eqnarray}
and
\begin{eqnarray}\label{e4deltafull}
&&f_{n,i}'+(2\sigma'+F)f_{n,i}=m_{n,i}e^{-\sigma}{\tilde f}_{n,i},\\ \label{e5deltafull}
&&{\tilde f}_{n,i}'+(2\sigma'-F){\tilde f}_{n,i}-\frac{K}{\beta_{i}}f_{n,i}\,\delta(z-L)=-m_{n,i}e^{-\sigma}f_{n,i},
\end{eqnarray}
which are valid for any $z$. Here $m_{0}\ge 0$, $m_{n,i}>0$. An interesting observation is that for $F(z)\equiv 0$ we can define the functions
$$
\hat f_{n,1}(z)=f_{n,1}(z)-{\tilde f}_{n,1}(z),\qquad \hat f_{n,2}(z)=f_{n,2}(z)+{\tilde f}_{n,2}(z),
$$
which satisfy the equations
\begin{eqnarray}\label{1dim1F0}
&&{\hat f}_{n,1}'(z)+2\sigma'(z){\hat f}_{n,1}(z)+K\,\delta(z-L){\hat f}_{n,1}(z)=m_{n,1}e^{-\sigma(z)}{\hat f}_{n,1}(-z),\\ \label{1dim2F0}
&&{\hat f}_{n,2}'(z)+2\sigma'(z){\hat f}_{n,2}(z)+K\,\delta(z-L){\hat f}_{n,2}(z)=-m_{n,2}e^{-\sigma(z)}{\hat f}_{n,2}(-z),
\end{eqnarray}
where we have used the symmetry properties of the functions $f_{n,i}(z)$, ${\tilde f}_{n,i}(z)$, equations (\ref{e4deltafull}), (\ref{e5deltafull}) and the regularization ${\tilde f}(z)\delta(z-L)=0$ in the derivation. These equations coincide with equations (\ref{1dim1}), (\ref{1dim2}) with $G(z)\equiv K\,\delta(z-L)$. This implies that most probably $\beta_{0}=\beta_{1}=1$, at least in the physically reasonable cases (it is also supported by the numerical calculations for the case $\sigma(z)\equiv 0$, $F(z)\sim\textrm{sign}(z)$).

Finally, substituting (\ref{dec1deltafull}), (\ref{dec2deltafull}) into five-dimensional action (\ref{fermactdeltaK}), using the orthogonality conditions (\ref{orthdelta}), (\ref{orthdeltamix}), equations (\ref{e4deltafullzero})--(\ref{e5deltafull}) and the normalization conditions
\begin{equation}\label{normdelta}
\int\limits_{-L}^{L}e^{3\sigma}\left(f_{0}^{2}(z)+{\tilde f}_{0}^{2}(z)\right)dz=1,\quad \int\limits_{-L}^{L}e^{3\sigma}\left(f_{n,i}^{2}(z)+{\tilde f}_{n,i}^{2}(z)\right)dz=1,\quad n\ge 1, i=1,2,
\end{equation}
we get\footnote{For the calculations, it is convenient to use the unified formula for the orthogonality conditions (\ref{orthdelta}), (\ref{orthdeltamix}) and the normalization conditions (\ref{normdelta}):
$\int_{-L}^{L}e^{3\sigma}\left(f_{n,i}(z)f_{k,j}(z)+\beta_{i}\beta_{j}{\tilde f}_{n,i}(z){\tilde f}_{k,j}(z)\right)dz=\int_{-L}^{L}e^{3\sigma}\left(\beta_{i}\beta_{j}f_{n,i}(z)f_{k,j}(z)+{\tilde f}_{n,i}(z){\tilde f}_{k,j}(z)\right)dz=\delta_{nk}\delta_{ij}$.}
\begin{eqnarray}
S_{eff}=\int d^{4}x\Biggl(i\bar\psi\gamma^{\mu}\partial_{\mu}\psi-m_{0}\bar\psi\psi+\sum\limits_{n=1}^{\infty}\sum\limits_{i=1}^{2}\left(
i\bar\psi_{n,i}\gamma^{\mu}\partial_{\mu}\psi_{n,i}-m_{n,i}\bar\psi_{n,i}\psi_{n,i}\right)\Biggr),
\end{eqnarray}
where $\psi=\psi_{L}^{0}+\psi_{R}^{0}$, $\psi_{n,i}=\psi_{L}^{n,i}+\psi_{R}^{n,i}$. It is necessary to point out that since we used the regularization ${\tilde f}(z)\delta(z-L)\sim\textrm{sign}(z-L)\delta(z-L)=0$ in the equations of motion in order to make them self-consistent, it should be used in the derivation of the effective action too, i.e., the term ${\tilde f}^{2}(z)\,\delta(z-L)$ should be handled as $${\tilde f}^{2}(z)\,\delta(z-L)={\tilde f}(z)\Bigl({\tilde f}(z)\delta(z-L)\Bigr)\sim {\tilde f}(z)\Bigl(\textrm{sign}(z-L)\delta(z-L)\Bigr)={\tilde f}(z)\times 0=0,$$ but not as $${\tilde f}^{2}(z)\delta(z-L)\sim \textrm{sign}^{2}(z-L)\delta(z-L)=\delta(z-L).$$
This point reflects the well-known fact that the algebra of generalized functions is not associative.

It is necessary to point out that:
\begin{enumerate}
\item Contrary to the previous cases, there are no second-order differential equations for the wave functions, which are valid for any $z$ --- the point $z=L$ is excluded. However, it becomes possible to construct the systems of first-order differential equations (\ref{e4deltafullzero})--(\ref{e5deltafull}), which are valid for any $z$. Note that these systems of equations are different for different modes (they depend on $\beta_{i}$).
\item The orthogonality conditions have a nonstandard form (\ref{orthdelta}) and (\ref{orthdeltamix}), which also depends on the type of the modes ($i=1$, $i=2$ or mixed).
\item Contrary to the case $F(z)\equiv 0$, here we can not replace the delta-function by its smooth approximation, for example, by the Gaussian profile. Such a change will break the structure of the equations of motion and will lead to fourth-order differential equations, at least if $F(z)$ is not fine-tuned as in (\ref{partic2}). In fact, here $G(z)\equiv K\,\delta(z-L)$ serves as a source of nonstandard boundary conditions for the five-dimensional fields.
\end{enumerate}

\subsection{Small discussion}
Let us briefly discuss the results presented above.
\begin{enumerate}
\item The general case. As was shown in Section~3, this case naively contains rather pathological behavior in the form of fourth-order differential equations of motion for the components of the five-dimensional spinor fields. In principle, such equations of motion indicate that there may (but not necessarily) appear tachyons or ghosts in the effective theory. But even if such pathologies are absent for fermions in five-dimensional brane world models, at the moment it is unclear how to perform the Kaluza-Klein decomposition in this case (or even to isolate the lowest mode in a mathematically consistent way) and what is the number of physical degrees of freedom at each Kaluza-Klein level. Indeed, in all the cases discussed in this section the independent second-order equations of motion for the components of the five-dimensional fields (or of their linear combinations) suggested the correct separation of variables and provided the complete set of possible physical degrees of freedom of the four-dimensional effective theory in a rather model-independent way (at least the forms of $\sigma(z)$, $F(z)$ were not specified). It is not clear how to perform an analogous analysis starting from the fourth-order equations (\ref{hder1}), (\ref{hder2}).

    Due to the nonperturbative nature of the origin of the fourth-order differential equations of motion, it seems that the standard perturbation analysis does not describe all the physical degrees of freedom, analogously to the much more simple cases of four-dimensional scalar fields \cite{Woodard:2006nt}. Anyway, this case calls for a detailed and thorough analysis.
\item The case $F(z)\not\equiv 0$, $G(z)\equiv Me^{-\sigma}$ does not contain any pathological behavior, the consistent Kaluza-Klein decomposition can also be performed in this case. A remarkable feature of this choice of the parameters is that the form of the zero mode wave function does not depend on the value of the four-dimensional mass of this mode.

    However, this case may contain some serious drawbacks. Indeed, the existence of the functions $\sigma(z)\not\equiv 0$, $F(z)\not\equiv 0$ and $G(z)\equiv Me^{-\sigma}$, whatever the origin of these functions is, imply that the extra dimension is not uniform in $z$. Thus, one can expect that the backreaction of the bulk fields on the background metric or possible quantum corrections are also nonuniform in $z$, which may violate the fine-tuned relation between the vacuum profile of the Higgs field and the form of the background metric. Although the violation can be very small, the effect is nonperturbative, so it will lead to the problems discussed in the previous item. Drawing an analogy with the four-dimensional example (\ref{4Dscact}), the case $F(z)\not\equiv 0$, $G(z)\equiv Me^{-\sigma}$ is similar to the case $\alpha=0$ in (\ref{4Dscact}): it gives the second-order equations of motion for the components of the fields $\Psi_{1}$ and $\Psi_{2}$ from the very beginning, but deviations from $G(z)\equiv Me^{-\sigma}$ may, in principle, lead to pathologies similar to those with $0<|\alpha|\ll 1$ in (\ref{4Dscact}). This point should be taken into account when considering this case.
\item The case $G(z)\equiv K\,\delta(z-L)$, $F(z)\not\equiv 0$ also admits a consistent Kaluza-Klein decomposition and does not contain any pathologies. However, it leads to the discontinuous wave functions of the modes. This discontinuity imply that such a wave function is just an approximation for a continuous, but very rapidly varying in the vicinity of $z=L$ wave function, whereas the delta-function-like $G(z)$ is an approximation of a very narrow and peaked, but continuous profile of the Higgs field.\footnote{In standard brane world models, the branes interact with gravitation as classical objects, so ``infinitely thin brane'' as a classical object is an idealization. This implies that delta-functions or step functions should also be considered as idealizations, whereas consistent theories should admit the replacements of generalized functions by regular functions.} However, any modification of the delta-function (such as, for example, $\delta(z-L)$ $\to$ Gaussian profile of $G(z)$) in this case leads to impossibility to diagonalize the matrices $\hat\Lambda_{L,R}$ in (\ref{HatLambdadef}) and, consequently, to fourth-order equations of motion for the components of the five-dimensional fermion fields. This situation is also similar in some sense to the case $\alpha=0$ in (\ref{4Dscact}).

    The cases (\ref{partic1}), (\ref{partic2}) seem to possess the same problems. Though it looks as if these cases allow for consistent Kaluza-Klein decompositions in principle, they are highly fine-tuned. First, it is impossible to pass to smooth functions $F(z)$ and $G(z)$ --- the existence of $\textrm{sign}(z)$ is necessary for the diagonalization of $\hat\Lambda_{L,R}$ in (\ref{HatLambdadef}). For example, for $F(z)\sim\textrm{sign}(z)$ and with $M=0$ the choice (\ref{partic1}) leads to the natural profile $G(z)\equiv\textrm{const}$. However, if we suppose that $\textrm{sign}(z)$ is an idealization for $F(z)$ and, in analogy with the original Rubakov-Shaposhnikov mechanism \cite{RS}, replace $\textrm{sign}(z)$ in $F(z)$ by $-\tanh(C(z-L))$ in the vicinity of $z=L$, where $C$ is a constant, we will either get the unnatural form $G(z)\sim \textrm{sign}(z)\tanh(C(z-L))$ for the vacuum profile of the Higgs field or violate the fine-tuned relation (\ref{partic1}), if we keep $G(z)\equiv\textrm{const}$.

    Second, as it was noted above, the backreaction of the bulk fields on the background metric or possible quantum corrections to the Higgs field potential can be nonuniform in $z$, which may modify $\sigma(z)$ and the vacuum profile of the Higgs field (i.e., $G(z)$) keeping $F(z)$ intact. The latter will violate the fine-tuned relations in (\ref{partic1}), (\ref{partic2}), again leading to the fourth-order equations of motion. Drawing an analogy with the four-dimensional example (\ref{4Dscact}), the cases (\ref{partic1}), (\ref{partic2}) are similar to the case $\alpha=-\frac{1}{4}$ in (\ref{4Dscact}).

\item Contrary to the previous cases, the choice $F(z)\equiv 0$ looks the most safe from this point of view (of course, if some quantum correction do not induce a nonzero effective correction to the zero value of $F(z)$). Indeed, the Kaluza-Klein decomposition can be performed in a mathematically consistent way for any $G(z)$ without the necessity for any fine-tuning. The chiral structure of the zero Kaluza-Klein mode for $(e^{\sigma}G)'\not\equiv 0$ appears to be more complicated than the one in the case $G(z)\equiv Me^{-\sigma}$. The wave function of the zero mode now depends on the mass of this mode, but for the physically reasonable cases these effects can be calculated perturbatively. Again drawing an analogy with the four-dimensional example (\ref{4Dscact}), the case $F(z)\equiv 0$ is similar to the case $\alpha=-\frac{1}{4}$ in (\ref{4Dscact}), but now it stays in this ``degenerate'' point with any $G(z)$ .
\end{enumerate}

\section{Reproducing the Standard Model by the zero Kaluza-Klein modes in the Randall-Sundrum background}
It is clear that in the flat background, discussed in Section~2 (i.e., in the simplest example of a model with universal extra dimensions \cite{Appelquist:2000nn}), the zero mode sector of the theory {\em exactly} reproduces the Standard Model. This happens because the wave functions of all the zero modes are just constants in that case.  But it is unclear whether such a possibility to exactly reproduce the Standard Model by the zero mode sector exists in the warped case.

The motivation for this study is the following. In the previous sections it was shown that there are several cases, for which the Kaluza-Klein decomposition for fermions can be performed consistently and in the correct mathematical way, leading to different forms of the wave function and chiral structure even of the zero Kaluza-Klein mode. However, it is well known that in the general case the interaction of gauge fields with the Higgs field results in a modification of the shapes of the zero mode gauge boson wave functions. The latter results in a modification of the coupling constants, leading to severe restrictions on the fundamental energy scale of five-dimensional theory \cite{Csaki:2002gy,Burdman:2002gr}, and these restrictions come mainly from the zero mode sector of the effective four-dimensional theory. So, there arises a question: which set of parameters allows one to reproduce at least the electroweak sector of the Standard Model by the zero Kaluza-Klein modes most closely without imposing restrictions on the size of the extra dimension?  Below we will examine this topic for the case of the Randall-Sundrum background metric \cite{Randall:1999ee}.

\subsection{Setup}
To start with, let us consider a five-dimensional action, describing fermion fields minimally coupled to the $SU(2)\times U(1)$ gauge fields in the Randall-Sundrum background with $\sigma(z)\equiv kL-k|z|$, of the form
\begin{eqnarray}\nonumber
S=\int d^{4}xdz\sqrt{g}\left(-\frac{\xi^2}{4}F^{a,MN}F^{a}_{MN}-\frac{\xi^2}{4}B^{MN}B_{MN}+g^{MN}\left(D_{M}H\right)^{\dag}D_{N}H-V(H^{\dag}H)\right.\\ \nonumber\left.+iE_{N}^{M}{\bar{\hat\Psi}_{1}}\Gamma^{N}D_{M}\hat\Psi_{1}+iE_{N}^{M}\bar\Psi_{2}\Gamma^{N}D_{M}\Psi_{2}-\sqrt{2}Y_{5}\left[\left({\bar{\hat\Psi}_{1}}
H\right)\Psi_2+\textrm{h.c.}\right]\right)\\ \label{action2}-\int_{z=0} d^{4}x\sqrt{g^{\textrm{ind}}}\left(+2kH^{\dag}H\right)-\int_{z=L} d^{4}x\sqrt{g^{\textrm{ind}}}\left(-2kH^{\dag}H+V_{br}(H^{\dag}H)\right),
\end{eqnarray}
where the factor $$\xi=\frac{1}{\sqrt{2L}}$$ is a constant, which is introduced for convenience and chosen so that the dimension of the bulk gauge fields is mass (in this case the coupling constants $g$ and $g'$ are dimensionless); $g^{\textrm{ind}}_{\mu\nu}$ is the induced metric on the branes;
\begin{equation}\label{Higgspot}
V(H^{\dag}H)=-3k^2H^{\dag}H,\qquad V_{br}(H^{\dag}H)=\lambda_{br}\left(H^{\dag}H-\frac{v_{br}^{2}}{2}\right)^2,
\end{equation}
where $V(H^{\dag}H)$ is the fine-tuned bulk scalar field potential \cite{Smolyakov:2015zsa} and $V_{br}(H^{\dag}H)$ with $\lambda_{br}>0$ is the scalar field potential on the brane which will provide a nonzero mass of the Higgs boson (also take into account the fine-tuned terms $\pm 2kH^{\dag}H$ on the branes in (\ref{action2}));
\begin{eqnarray}\label{F1}
F^{a}_{MN}&=&\partial_{M}A^{a}_{N}-\partial_{N}A^{a}_{M}+g\epsilon^{abc}A^{b}_{M}A^{c}_{N},\\ \label{F1a}
B_{MN}&=&\partial_{M}B_{N}-\partial_{N}B_{M},\\ \label{F2}
D_{M}H&=&\left(\partial_{M}-ig\frac{\tau^{a}}{2}A^{a}_{M}-i\frac{g'}{2}B_{M}\right)H
\end{eqnarray}
and the fields satisfy the orbifold symmetry conditions $A^{a}_{\mu}(x,-z)=A^{a}_{\mu}(x,z)$, $A^{a}_{5}(x,-z)=-A^{a}_{5}(x,z)$, $B_{\mu}(x,-z)=B_{\mu}(x,z)$, $B_{5}(x,-z)=-B_{5}(x,z)$, $H(x,-z)=H(x,z)$. In what follows, we will use the gauge $A^{a}_{5}(x,z)\equiv 0$, $B_{5}(x,z)\equiv 0$. We use the extra constant $kL$ in the Randall-Sundrum solution for $\sigma(z)$ just for convenience, in order to have Galilean coordinates on the brane at $z=L$ and to refer the energy units to these coordinates \cite{Rubakov,Boos:2002hf}.

The $SU(2)$ doublet, constructed from five-dimensional spinors, is denoted by
\begin{eqnarray}
{\hat\Psi}_{1}=\left(
\begin{array}{l}
\Psi^{\nu} \\ \Psi_{1}\\
\end{array}
\right),
\quad {\bar{\hat\Psi}_{1}}=\left(\bar\Psi^{\nu}\,,\,\bar
\Psi_{1}\right)
\end{eqnarray}
and the five-dimensional $SU(2)$ singlet is denoted by $\Psi_{2}$. The covariant derivatives are
defined by
\begin{eqnarray}
D_{M}{\hat\Psi}_{1}&=&\left(\nabla_{M}-ig\frac{\tau^{a}}{2}A_{M}^{a}+i\frac{g'}{2}B_{M}\right){\hat\Psi}_{1},\\
D_{M}\Psi_{2}&=&\left(\nabla_{M}+ig'B_{M}\right)\Psi_{2}.
\end{eqnarray}
It is clear that the first two lines of action (\ref{action2}) describe just the five-dimensional generalization of the electroweak sector of the Standard Model with one generation of leptons.

The remarkable feature of the scalar field potential in (\ref{action2}) is that it provides a special form of the vacuum solution for the Higgs field. Indeed, the vacuum solution, breaking the gauge group $SU(2)\times U(1)$ to $U(1)_{em}$, leaving the Poincare invariance in four-dimensional space-time intact and satisfying the corresponding equations of motion, takes the form
\begin{equation}\label{vev2}
H_{0}(z)\equiv\frac{1}{\sqrt{2}}\left(0\atop v_{br}e^{-\sigma}\right)\equiv\frac{1}{\sqrt{2}}\left(0\atop v_{br}e^{k|z|-kL}\right).
\end{equation}
The constant $v_{br}$ in front of the term $e^{k|z|-kL}$ in (\ref{vev2}) is fixed by the brane localized potential $V_{br}(H^{\dag}H)$, see (\ref{Higgspot}); in the absence of this potential there would be an arbitrary constant instead of $v_{br}$. It will be shown below that, though the bulk Higgs field potential is not bounded from below, the spectrum of Kaluza-Klein modes does not contain tachyons, indicating that the vacuum solution is classically stable under small perturbations. All the other fields are identically zero in the vacuum. The backreaction of the Higgs vacuum field on the background metric is also neglected (its validity will be checked later).

The zero Kaluza-Klein modes of this theory are supposed to reproduce the Standard Model fields, so below we will focus only on the zero modes of the theory and neglect all the higher Kaluza-Klein modes of the gauge, fermion and Higgs fields. We set the localization function $F(z)\equiv 0$ in order not to worry about possible small corrections (such as backreaction of the fields on the background metric or quantum corrections), which could lead to nonperturbative effects and possible pathologies, --- as follows from the results of Section~4, in the case $F(z)\equiv 0$ such a violation of the fine-tuned relation between the profile of the Higgs field and the form of the background metric may lead only to a small modification of the chiral structure of fermions, as well as to a small modification of the gauge boson wave functions.

\subsection{Gauge boson and fermion zero modes}
From the very beginning, with the help of transformations
\begin{eqnarray}\label{gaugephys}
Z_{\mu}=\frac{1}{\sqrt{g^2+{g'}^2}}\left(gA^{3}_{\mu}-g'B_{\mu}\right),\,\, A_{\mu}=\frac{1}{\sqrt{g^2+{g'}^2}}\left(g B_{\mu}+g'A^{3}_{\mu}\right),\,\,
W^{\pm}_{\mu}=\frac{1}{\sqrt{2}}\left(A^{1}_{\mu}\mp iA^{2}_{\mu}\right),
\end{eqnarray}
it is convenient to pass  to the physical degrees of freedom of the theory. It is not difficult to show that the equations for the wave functions and the masses of the Kaluza-Klein modes in the vacuum (\ref{vev2}) are just \cite{Smolyakov:2015zsa}
\begin{eqnarray}\label{Wwf}
-m_{W,n}^{2}f_{W,n}-\partial_{5}(e^{2\sigma}\partial_{5}f_{W,n})+\frac{g^2}{4\xi^2}v_{br}^2f_{W,n}=0,\\ \label{Zwf}
-m_{Z,n}^{2}f_{Z,n}-\partial_{5}(e^{2\sigma}\partial_{5}f_{Z,n})+\frac{g^2+{g'}^{2}}{4\xi^2}v_{br}^2f_{Z,n}=0,\\ \label{Awf}
-m_{A,n}^{2}f_{A,n}-\partial_{5}(e^{2\sigma}\partial_{5}f_{A,n})=0.
\end{eqnarray}
Remarkably, for all the zero modes (from here and below we omit the superscript ``0'' for the zero ($n=0$) modes of the fields) the normalized wave functions are (recall the factor $\xi=\frac{1}{\sqrt{2L}}$ in (\ref{action2}))
\begin{equation}\label{GBwf}
f_{W}(z)\equiv 1, \qquad f_{Z}(z)\equiv 1, \qquad f_{A}(z)\equiv 1,
\end{equation}
and
\begin{equation}\label{gmasses}
m_{W}=\frac{g v}{2},\qquad
m_{Z}=\frac{\sqrt{g^2+{g'}^{2}} v}{2},\qquad m_{A}=0,
\end{equation}
where
\begin{equation}\label{tildevdef}
v=\frac{v_{br}}{\xi}=v_{br}\sqrt{2L}
\end{equation}
must be identified with the vacuum expectation value of the Higgs field in the Standard Model.

Situation with the fermions is also very simple. Indeed, according to (\ref{zm1}), (\ref{zm2}) with $F(z)\equiv 0$, the zero modes of the fermion fields can be represented as
\begin{eqnarray}\label{substferm}
{\hat\Psi}_{1}(x,z)=Ce^{-2\sigma(z)}\left(
\begin{array}{l}
\nu_{L}(x) \\ \psi_{L}(x)\\
\end{array}\right),\quad \Psi_{2}(x,z)=Ce^{-2\sigma(z)}\psi_{R}(x),
\end{eqnarray}
where $\frac{1}{C^{2}}=\int\limits_{-L}^{L}e^{-\sigma}dz$ is a normalization constant. We see that all the fermion zero modes have the same wave function $f(z)=Ce^{-2\sigma(z)}$. According to (\ref{zm1}), (\ref{zm2}), (\ref{vev2}) and (\ref{action2}), the masses of the zero modes are
\begin{eqnarray}\label{zmmasspsi}
m_{\psi}=Y_{5}v_{br}=Yv,\\
m_{\nu}=0,
\end{eqnarray}
where $Y=Y_{5}\xi=\frac{Y_{5}}{\sqrt{2L}}$.

Substituting the solutions for the zero modes of the gauge and fermion fields into the five-dimensional action and integrating over the coordinate of the extra dimension, we get exactly the gauge boson and lepton sectors of the Standard Model. We do not present here the calculations and the result, they are straightforward. This happens because the overlap integrals, involving the wave functions of fermions and gauge bosons reduce to $$\int e^{3\sigma}f^{2}(z)dz=1$$ (because of the fact that the wave functions of gauge bosons are just constants, whereas the wave functions of all fermion fields are the same). An interesting observation is that the terms describing the self-interaction of gauge bosons also appear to be exactly the same as those in the Standard Model --- again this happens due to the fact that the wave functions of gauge bosons are just constants. It should also be noted that two more generations of leptons, as well as quarks and gluons, can be added to the theory in a fully analogous way.

At this step we can make the observation that in the case under consideration there is no modification of the gauge boson wave function due to the interaction with the vacuum solution of the Higgs field. The constant wave functions of gauge bosons (\ref{GBwf}) ensure the charge universality \cite{Rubakov} and allow one not to worry about the problems caused by the modification of the gauge boson wave functions and its effect on the precision electroweak data, discussed in \cite{Csaki:2002gy,Burdman:2002gr}. Of course, the higher Kaluza-Klein modes still affect the four-dimensional effective theory and their contribution puts constraints on the fundamental parameters of the theory. The corresponding calculations are very complicated and should take into account the contributions coming from the Kaluza-Klein modes of the gauge, fermion and Higgs fields, including the nontrivial interactions between them. But this issue lies beyond the scope of the present paper and such calculations will not be presented here.

Thus, it is possible to exactly reproduce the fermion and gauge boson sectors of the Standard Model by the zero Kaluza-Klein modes in the five-dimensional warped brane world model. Note that though we take the Randall-Sundrum background metric \cite{Randall:1999ee} as the metric of the model under consideration, its explicit form has not been used for deriving the effective action for the gauge boson and fermion zero modes --- it was obtained in a model-independent way using the results of Section~4.\footnote{It is interesting to note that analogous results can be obtained for the case $F(z)\not\equiv 0$ (but, of course, with $G(z)\equiv Me^{-\sigma}$) too, but, as it was noted above, such a case is not protected from the nonperturbative effects discussed above.} However, as we will see below, in the Higgs sector there are some small deviations from the Standard Model already in the case of the Randall-Sundrum background. Their calculation demands the explicit form of the metric and wave functions of the fields. So, let us proceed to the Higgs field.

\subsection{The Higgs field}
First, let us check that the backreaction of the vacuum solution of the Higgs field on the background metric can be neglected. To do it, we will simply compare the value of the five-dimensional cosmological constant $\Lambda=-24M_{5}^{3}k^{2}$ \cite{Randall:1999ee}, where $M_{5}$ is the five-dimensional Planck mass, with the values of the bulk scalar field potential $V(H^{\dag}H)=-3k^2H^{\dag}H$ in the vacuum (\ref{vev2}). For simplicity, we also suppose that $M_{5}\approx k$ and thus $|\Lambda|\approx 24k^{5}$. Then, we get
\begin{eqnarray}
|V(H^{\dag}H)|=3k^2H^{\dag}H=\frac{3}{2}k^{2}e^{-2\sigma}v_{br}^{2}\le \frac{3}{2}k^{2}v_{br}^{2}=\frac{3}{2}k^{2}\frac{v^{2}}{2L}=\frac{3}{4kL}\frac{v^{2}}{k^{2}}k^{5}\approx\frac{1}{32kL}\frac{v^{2}}{k^{2}}|\Lambda|,
\end{eqnarray}
where we have used (\ref{tildevdef}). For $v=246\,GeV$, $k=2\,TeV$ and $kL=35$ we get
\begin{eqnarray}
|V(H^{\dag}H)|\lesssim 1.35\times 10^{-5}|\Lambda|\ll|\Lambda|.
\end{eqnarray}
The latter shows that the Randall-Sundrum background metric remains intact with a good accuracy under the influence of the vacuum solution of the Higgs field.

Now we represent the Higgs field $H$ as
\begin{equation}\label{higgsfield}
H(x,z)\equiv e^{-\sigma}\left(\rho_{1}(x,z)+i\rho_{2}(x,z)\atop \frac{1}{\sqrt{2}}\left(v_{br}+\chi(x,z)\right)+i\rho_{3}(x,z)\right),
\end{equation}
where $\chi$, $\rho_{1}$, $\rho_{2}$ and $\rho_{3}$ are real fields. Substituting this representation into the equation of motion, coming from (\ref{action2}), and retaining only the linear terms in the fields $\chi$, $\rho_{1}$, $\rho_{2}$ and $\rho_{3}$, we get
\begin{eqnarray}\label{chieq}
\left(e^{2\sigma}\chi'\right)'-\Box\chi-2\lambda_{br} v_{br}^{2}\delta(z-L)\chi&=&0,\\ \label{xieq}
\left(e^{2\sigma}\rho_{i}'\right)'-\Box\rho_{i}&=&0,\quad i=1,2,3.
\end{eqnarray}
It is not difficult to show that the Kaluza-Klein masses of the fields $\chi$, $\rho_{1}$, $\rho_{2}$ and $\rho_{3}$ are real (see Appendix~E), i.e., there are no tachyons in the spectra of the modes, though the bulk scalar field potential in unbounded from below (\ref{Higgspot}). Below we will focus only on the zero modes.

First, it is clear that the solutions to equation (\ref{xieq}), which are connected with the zero mode wave functions of the fields $\rho_{i}$, $i=1,2,3$, are just constants, whereas the masses of these modes are $m_{\rho_{i},0}=0$. Due to this fact, using the residual gauge transformations, which are left after imposing the gauge $A^{a}_{5}(x,z)\equiv 0$, $B_{5}(x,z)\equiv 0$, we can set all the zero modes of the fields $\rho_{i}$, $i=1,2,3$ identically to zero. This can be done exactly in the same way as in the four-dimensional Standard Model, so we will not discuss this issue in detail. It should be noted that higher Kaluza-Klein modes of these fields can not be gauged out and should be taken into account in the corresponding calculations.

Thus, the unitary gauge can be imposed on the zero mode sector of the Higgs field. The equation for the reduced wave function of the zero mode $f_{\chi,0}(z)=f_{h}(z)$ of the field $\chi$ (the entire wave function is $e^{-\sigma}f_{h}(z)$, see (\ref{higgsfield})) takes the form
\begin{equation}\label{chieqzm}
\left(e^{2\sigma}f_{h}'(z)\right)'+m_{h}^{2}f_{h}(z)-2\lambda_{br} v_{br}^{2}\delta(z-L)f_{h}(z)=0,
\end{equation}
where $m_{h}=m_{\chi,0}$ is the mass of the zero mode. In what follows, we will consider the term with $\lambda_{br}$ as a perturbation and represent $f_{h}(z)$ as
\begin{equation}\label{fdefthroughg}
f_{h}(z)=\frac{1}{\sqrt{2L}}(1+g_{h}(z)).
\end{equation}
For $g_{h}(z)$ in the leading order we obtain
\begin{equation}\label{chieqzm1}
\left(e^{2\sigma}g_{h}'(z)\right)'+m_{h}^{2}-2\lambda_{br} v_{br}^{2}\delta(z-L)=0.
\end{equation}
Integrating this equation over the coordinate of the extra dimension $z$, we obtain
\begin{equation}\label{HiggsmassSM}
m_{h}^2=\frac{2\lambda_{br} v_{br}^{2}}{2L}=2\lambda v^{2},
\end{equation}
where $v$ is defined by (\ref{tildevdef}) and
\begin{equation}\label{lambdaSM}
\lambda=\frac{\lambda_{br}}{4L^{2}}=\lambda_{br}\xi^{4}.
\end{equation}
Solving equation (\ref{chieqzm1}), we can get the properly normalized (up to and including the terms $\sim\frac{m_{h}^2}{k^{2}}$) approximate solution for $f_{h}(z)$, which has the form
\begin{eqnarray}\nonumber
f_{h}(z)=\frac{1}{\sqrt{2L}}(1+g_{h}(z))\approx\frac{1}{\sqrt{2L}}\left(1+\frac{m_{h}^2}{4k^{2}}\left(\frac{kL-1+e^{-2kL}}{kL}-(2k|z|-1)e^{2k|z|-2kL}\right)\right)\\ \label{solpertfh} \approx \frac{1}{\sqrt{2L}}\left(1+\frac{m_{h}^2}{4k^{2}}\left(\frac{kL-1}{kL}-(2k|z|-1)e^{2k|z|-2kL}\right)\right).
\end{eqnarray}
It is clear that if $\lambda_{br}=0$ (i.e., if we turn off the brane scalar field potential), the mass of the Higgs boson is equal to zero, whereas the wave function of the Higgs boson is proportional to the vacuum profile of the Higgs field $e^{-\sigma}$. Since for solution (\ref{solpertfh}) the relation
\begin{equation}\label{intfh1}
\int\limits_{-L}^{L}g_{h}(z)dz=0
\end{equation}
holds, the normalization condition
\begin{equation}\label{intfh2}
\int\limits_{-L}^{L}f_{h}^{2}(z)dz=1
\end{equation}
is fulfilled up to and including the terms of the order of $\frac{m_{h}^2}{k^{2}}$.

An important remark is in order here. The perturbation analysis and solution (\ref{solpertfh}) for the Higgs boson wave function make sense only if
$$|g_{h}(z)|\ll 1.$$
It is not difficult to find that the maximum of $|g_{h}(z)|$ is attained at $z=L$, $|g_{h}(L)|\approx 17\,\frac{(125\,GeV)^2}{k^{2}}$, where we have used $kL=35$ and $m_{h}=125\,GeV$. We will restrict ourselves to considering the values of the parameter $k$ such that
\begin{equation}
k\gtrsim 2\,TeV,
\end{equation}
for which
\begin{equation}
|g_{h}(L)|\lesssim 0.066,
\end{equation}
which looks rather reasonable. For the smaller values of $k$ one should obtain an exact solution for the Higgs boson wave function.

Now we are ready to calculate the couplings of the zero mode of the Higgs field
\begin{equation}\label{chi0}
\chi_{0}(x,z)=f_{h}(z)h(x),
\end{equation}
in which the four-dimensional field $h(x)$ can be identified with the Standard Model Higgs boson. We will calculate all the couplings to gauge bosons and fermions up to and including the terms of the order of $\frac{m_{h}^2}{k^{2}}$.

First, substituting (\ref{chi0}) into the five-dimensional action of the Higgs field, integrating over the coordinate of the extra dimension and using (\ref{chieqzm}), (\ref{lambdaSM}) and (\ref{tildevdef}), we get the standard action resembling the one of the Standard Model Higgs boson
\begin{equation}\label{higgsbosact}
\int d^{4}x\left(\frac{1}{2}\partial^{\mu}h\partial_{\mu}h-\frac{m_{h}^{2}}{2}h^{2}+(1+\delta_{3})\lambda v h^{3}+(1+\delta_{4})\frac{\lambda}{4}h^{4}\right),
\end{equation}
where
\begin{equation}
\delta_{3}=\left(\sqrt{2L}\,f_{h}(L)\right)^{3}-1,\qquad \delta_{4}=\left(\sqrt{2L}\,f_{h}(L)\right)^{4}-1.
\end{equation}
For (\ref{HiggsmassSM}), (\ref{solpertfh}) and with $kL=35$ we have $\lambda v=\frac{m_{h}^{2}}{2v}$, $\lambda=\frac{m_{h}^{2}}{2v^{2}}$ and
\begin{eqnarray}
\delta_{3}\approx 3g_{h}(L)\approx-\frac{3m_{h}^{2}}{2k^{2}}\left(kL-1+\frac{1}{2kL}\right)\approx -51\frac{m_{h}^{2}}{k^{2}},\\
\delta_{4}\approx 4g_{h}(L)\approx-\frac{2m_{h}^{2}}{k^{2}}\left(kL-1+\frac{1}{2kL}\right)\approx -68\frac{m_{h}^{2}}{k^{2}}.
\end{eqnarray}
Although the brane scalar field potential with $\lambda_{br}$ (more precisely, the ``mass'' term which comes from this potential) is considered here as a perturbation, the self-coupling constants of the Higgs boson were calculated including the corrections of the order of $\frac{m_{h}^{2}}{k^{2}}$.

Now let us consider the interactions of the Higgs boson with the gauge bosons. In fact, all the interaction terms, in addition to the structures inherent to the Standard Model interaction terms, contain the overlap integrals of the form
\begin{equation}\label{overlap1}
\int\limits_{-L}^{L}e^{4\sigma}e^{-2\sigma}f_{W,Z}^{2}(z)(e^{-\sigma}v_{br})(e^{-\sigma}f_{h}(z))dz
\end{equation}
for $hWW$ and $hZZ$ interaction terms, and
\begin{equation}\label{overlap2}
\int\limits_{-L}^{L}e^{4\sigma}e^{-2\sigma}f_{W,Z}^{2}(z)(e^{-\sigma}f_{h}(z))^{2}(z)dz
\end{equation}
for $hhWW$ and $hhZZ$ interaction terms. The term $e^{4\sigma}$ comes from $\sqrt{g}$, whereas the term $e^{-2\sigma}$ comes from $g^{\mu\nu}$. Integral (\ref{overlap2}) is equal to unity regardless of the use of the perturbation theory --- with (\ref{GBwf}) it just comes to the normalization condition for the Higgs boson wave function. As for integral (\ref{overlap1}), using equations (\ref{GBwf}), (\ref{fdefthroughg}) and (\ref{intfh1}) we get
\begin{equation}\label{overlap1a}
\int\limits_{-L}^{L}e^{4\sigma}e^{-2\sigma}f_{W,Z}^{2}(z)(e^{-\sigma}v_{br})(e^{-\sigma}f_{h}(z))dz=
v_{br}\int\limits_{-L}^{L}f_{h}(z)dz=\sqrt{2L}v_{br}=v.
\end{equation}
Thus, the coupling constants of the Higgs boson to the gauge bosons appear to be the same as in the Standard Model (recall that $v$ should be identified with the Standard Model Higgs field vacuum expectation value), at least up to and including the terms of the order of $\frac{m_{h}^{2}}{k^{2}}$ for $hWW$ and $hZZ$ interactions.

The last step is to consider the interaction of the Higgs boson with fermions. The corresponding coupling constant to the field $\psi$ takes the form
\begin{eqnarray}\nonumber
\sqrt{2}Y_{5}\int\limits_{-L}^{L}e^{4\sigma}\left(Ce^{-2\sigma}\right)^{2}\left(\frac{1}{\sqrt{2}}e^{-\sigma}f_{h}(z)\right)dz=
\frac{Y_{5}}{\sqrt{2L}}\int\limits_{-L}^{L}e^{-\sigma}C^{2}\left(1+g_{h}(z)\right)dz\\\label{deltapsi}=
Y\left(1+\left(\int\limits_{-L}^{L}e^{-\sigma}dz\right)^{-1}\int\limits_{-L}^{L}e^{-\sigma}g_{h}dz\right)=Y\left(1+\delta_{\psi}\right),
\end{eqnarray}
where $Y=\frac{m_{\psi}}{v}$ is the Standard Model coupling constant. The integrals in (\ref{deltapsi}) can be easily evaluated, revealing
\begin{equation}
\delta_{\psi}\approx\frac{m_{h}^{2}}{4k^{2}}\left(\frac{14-6kL-\frac{9}{kL}}{9}\right)\approx -5.45\frac{m_{h}^{2}}{k^{2}}
\end{equation}
for $kL=35$, where we have dropped the terms $\sim e^{-kL}$ and smaller terms.

\subsection{Small discussion}
In this Section, an attempt was made to construct a five-dimensional theory in the Randall-Sundrum background with all the fields living in the bulk, such that its zero mode sector reproduces the Standard Model (namely, its electroweak and Higgs sectors) most closely. The results presented above show that the fermion and gauge boson sectors of the Standard Model can be exactly reproduced, including the interaction terms, by the zero Kaluza-Klein modes of the corresponding five-dimensional fields. This became possible because of the special choice of the fine-tuned bulk and brane potentials for the Higgs field, providing the necessary vacuum profile of the Higgs field. Although the analysis was performed using only one generation of leptons, two more generations, as well as the quark and gluon sectors, can be added to the theory (\ref{action2}) in an analogous way, leading to the same results.

The difference with the Standard Model arises in the interaction terms with the Higgs boson. The coupling constants to the gauge bosons appear to be the same as in the Standard Model (at least up to and including the terms of the order of $\frac{m_{h}^{2}}{k^{2}}$), but the coupling constants to fermions and self-coupling of the Higgs boson differ from those in the Standard Model. The relative deviations in these coupling constants can be encoded in the dimensionless parameters $\delta_{3}$ and $\delta_{4}$ for the self-couplings of the Higgs boson (\ref{higgsbosact}) and in the dimensionless parameter $\delta_{\psi}$ for the couplings to fermions (\ref{deltapsi}), such that
\begin{equation}
\delta_{3}\approx -51\frac{m_{h}^{2}}{k^{2}},\qquad \delta_{4}\approx -68\frac{m_{h}^{2}}{k^{2}},\qquad \delta_{\psi}\approx -5.45\frac{m_{h}^{2}}{k^{2}}.
\end{equation}
For example, for $k=2\,TeV$ and $m_{h}=125\,GeV$ we get
\begin{equation}
\delta_{3}\approx -0.2,\qquad \delta_{4}\approx -0.27,\qquad \delta_{\psi}\approx -0.02.
\end{equation}

For the values of the parameter $k<2\,TeV$ one may expect that, in addition to the deviations from the Standard Model in the self-couplings of the Higgs boson (\ref{higgsbosact}) and in the Higgs boson couplings to fermions, there will arise analogous deviations in the coupling constants of $hWW$ and $hZZ$ interactions. Indeed, the integral (\ref{overlap1})
\begin{equation}
v_{br}\int\limits_{-L}^{L}f_{h}(z)dz\neq v
\end{equation}
in the general case.

Of course, the higher Kaluza-Klein modes of the fields also affect the four-dimensional effective theory and their contribution should be taken into account. However, as it was noted above, such a complicated analysis lies beyond the scope of the present paper.

\section{Conclusion and final remarks}
In the present paper the Kaluza-Klein decomposition for the fermion fields living in the bulk of five-dimensional brane world models with compact extra dimension is examined in detail in a mathematically consistent way. The key feature of the analysis is the derivation of systems of first-order differential equations for the wave functions of the Kaluza-Klein modes of the fields, which allow one to obtain the four-dimensional effective action in a model-independent way. An important point is that in order to properly use the systems of first-order equations, it is necessary to have second-order differential equations for the components of the five-dimensional fermion fields (or of their linear combinations), which suggest the appropriate separation of variables and provide the complete set of possible physical degrees of freedom of the four-dimensional effective theory. It is shown that such second-order equations of motion can be obtained not in all the cases --- for the majority of five-dimensional fermion field Lagrangians, most of which are widely discussed in the literature for phenomenological reasons, the only obvious possibility is to get fourth-order differential equations for the components of five-dimensional fermion fields. Since these components of the five-dimensional spinors make up four-dimensional fermion fields, whereas higher-derivative theories are known to contain pathologies \cite{Woodard:2006nt}, this makes an obvious problem. More precisely, from this point of view the most of the cases, naively admitting a localization of the fermion zero mode at one of the branes and a generation of its mass, are disfavored. Of course, it is possible that there are some ways to solve the problem and to avoid fourth-order differential equations of motion or to solve them, which are not clear for me at the moment. However, I think that this problem should at least be mentioned, whereas these ``pathological'' cases deserve careful and thorough examination.

For some of the cases, for which the second-order differential equations for the wave functions can be obtained, the detailed Kaluza-Klein decomposition procedures, providing all the physical degrees of freedom of the corresponding four-dimensional effective theories, are presented and discussed in detail. It was found that the procedures of the Kaluza-Klein decomposition are completely different for different cases.

Using the general results, obtained in the paper, a special fine-tuned case was considered in order to examine the possibility to reproduce the ordinary four-dimensional Standard Model, including all the interactions of the fields, by the zero Kaluza-Klein modes most closely regardless of the size of the extra dimension (or, equivalently, the value of the five-dimensional energy scale) and without taking into account the higher Kaluza-Klein modes. As a particular background, the Randall-Sundrum solution for the metric was considered. It was shown that, with a special choice of the bulk and brane Higgs field potentials, it is possible to exactly reproduce the fermion and gauge boson sectors of the Standard Model including the interactions between the fields. However, the deviations from the Standard Model can not be fully avoided: the coupling constants of the Higgs boson to fermions, the Higgs boson self-coupling constants and, when the perturbation theory can not be used, the coupling constants of $hWW$ and $hZZ$ interactions differ from those of the Standard Model. In the case, when the perturbation theory can be used (roughly speaking, when the inverse anti-de Sitter radius $k\gtrsim 2\,TeV$), the deviations were calculated explicitly and were shown to be proportional to the ratio $\frac{(125\,GeV)^{2}}{k^{2}}$. However, one should bear in mind that the proposed model has a drawback --- there is a fine-tuning not only between the bulk and brane scalar field potentials, but also between the scalar field potentials and the five-dimensional background metric (through the parameter $k$). The latter looks rather unnatural, at least in the absence of a symmetry which can ensure such a fine-tuning.

\section*{Acknowledgements}
The author is grateful to E.~Boos and I.~Volobuev for useful discussions. The work was supported by the grant 14-12-00363 of the Russian Science Foundation.

\section*{Appendix~A: Relation between the first-order and second-order differential equations for the case $G(z)\equiv Me^{-\sigma}$}
Let us take equation (\ref{e4}) and differentiate it with respect to $z$. We get
\begin{eqnarray}\label{e2appendix}
f_n''+(2\sigma''+F')f_n+5\sigma'f_n'+\sigma'(m_n+M)e^{-\sigma}{\tilde f_n}+(F-3\sigma')f_n'-(m_n+M)e^{-\sigma}{\tilde f_n}'=0,
\end{eqnarray}
where we have used $2\sigma'=5\sigma'-3\sigma'$. After substituting $f_n'$ and ${\tilde f_n}'$ from equations (\ref{e4}) and (\ref{e5}) into the last two terms of (\ref{e2appendix}), all the terms with ${\tilde f_n}$ vanish and we arrive exactly at (\ref{e2}).

A fully analogous procedure can be performed with equation (\ref{e5}), resulting in (\ref{e3}).

\section*{Appendix~B: The absence of tachyonic modes in the case $F(z)\equiv 0$}
It is convenient to represent equation (\ref{secorderG}) as
\begin{equation}\label{secorderGappendix}
m^2f+e^{\sigma}\left(\partial_{5}+2\sigma'\right)e^{\sigma}\left(\partial_{5}+2\sigma'\right)f+e^{\sigma}\left(e^{\sigma}G\right)'f-e^{2\sigma}G^{2}f=0.
\end{equation}
Multiplying this equation by $e^{3\sigma}f$, integrating over the coordinate of the extra dimension $z$, performing the integration by parts in the two terms and combining the resulting terms, we arrive to the following equality:
\begin{equation}
m^2\int e^{3\sigma}f^{2}dz=\int e^{5\sigma}\left(f'+2\sigma'f+Gf\right)^{2}dz.
\end{equation}
Since both integrals are nonnegative, we get $m^2\ge 0$.

\section*{Appendix~C: The absence of tachyonic modes in the case $G(z)\equiv K\,\delta(z-L)$}
In order to examine a possible existence of tachyonic modes, let us change $m^{2}\to-\mu^{2}$ in equations (\ref{e2delta}) and (\ref{e3delta}), where $\mu^{2}>0$. In this case, instead of equations (\ref{e4delta}) and (\ref{e5delta}) we get
\begin{eqnarray}\label{e4deltatach}
&f'(z)+(2\sigma'+F)f(z)=\hat\beta\mu e^{-\sigma}{\tilde f}(z),\\
\label{e5deltatach} &{\tilde f}'(z)+(2\sigma'-F){\tilde f}(z)=\hat\beta\mu e^{-\sigma}f(z),
\end{eqnarray}
where $\hat\beta=1$ or $\hat\beta=-1$. Without loss of generality we can take $\hat\beta=1$. Substituting (\ref{dec1delta}), (\ref{dec2delta}) into equations (\ref{feqs1}), (\ref{feqs2}) with $G(z)\equiv K\delta(z-L)$ and using (\ref{e4deltatach}), (\ref{e5deltatach}), we get
\begin{eqnarray}\label{4Ddiracdelta1tach}
i\gamma^{\mu}\partial_{\mu}\psi_{L}+\mu\psi_{R}=0,\qquad
i\gamma^{\mu}\partial_{\mu}\psi_{R}-\mu\psi_{L}=0,\\ \label{4Ddiracdelta2tach}
i\gamma^{\mu}\partial_{\mu}\hat\psi_{L}-\mu\hat\psi_{R}=0,\qquad
i\gamma^{\mu}\partial_{\mu}\hat\psi_{R}+\mu\hat\psi_{L}=0
\end{eqnarray}
everywhere except $z=L$. At the point $z=L$ equations (\ref{feqs1}) and (\ref{feqs2}) again give (\ref{bound1delta}) and (\ref{bound2delta}). Substituting (\ref{bound1delta}) and (\ref{bound2delta}) into (\ref{4Ddiracdelta2tach}), we can find that (\ref{4Ddiracdelta1tach}) and (\ref{4Ddiracdelta2tach}) are consistent if the condition
\begin{equation}\label{bounddeltatach}
\lim\limits_{\epsilon\to +0}\frac{K^{2}f^{2}(L)}{4{\tilde f^{2}(L-\epsilon)}}=-1.
\end{equation}
holds. Clearly, this equation does not provide any roots.

\section*{Appendix~D: Orthogonality of the modes in the case $G(z)\equiv K\,\delta(z-L)$}
Let us take the integral
\begin{equation}
m_{n,i}\int\limits_{-L}^{L}e^{3\sigma}\left(f_{n,i}(z)f_{k,i}(z)+{\tilde f}_{n,i}(z){\tilde f}_{k,i}(z)\right)dz,
\end{equation}
where $i=1$ or $i=2$. Substituting equations (\ref{e4delta}), (\ref{e5delta}) say, for the fields $f_{n,i}$, ${\tilde f}_{n,i}$, into this integral, performing the integration by parts and again using equations (\ref{e4delta}), (\ref{e5delta}), we get
\begin{equation}\label{orthapp}
(m_{n,i}-m_{k,i})\int\limits_{-L}^{L}e^{3\sigma}\left(f_{n,i}(z)f_{k,i}(z)+{\tilde f}_{n,i}(z){\tilde f}_{k,i}(z)\right)dz=2e^{4\sigma(L)}\Bigl(f_{n,i}(L){\tilde f}_{k,i}(L)-{\tilde f}_{n,i}(L)f_{k,i}(L)\Bigr),
\end{equation}
where ${\tilde f}_{n,i}(L)=\lim\limits_{\epsilon\to+0}{\tilde f}_{n,i}(L-\epsilon)$, ${\tilde f}_{k,i}(L)=\lim\limits_{\epsilon\to+0}{\tilde f}_{k,i}(L-\epsilon)$ and the symmetry properties of the functions $f_{n,i}(z)$, $f_{k,i}(z)$, ${\tilde f}_{n,i}(z)$, ${\tilde f}_{k,i}(z)$ were taken into account. With (\ref{bounddeltaextra}), the r.h.s. of equation (\ref{orthapp}) can be represented as
$$
2e^{4\sigma(L)}\Bigl(f_{n,i}(L){\tilde f}_{k,i}(L)+{\tilde f}_{n,i}(L)f_{k,i}(L)\Bigr)=-Ke^{4\sigma(L)}f_{n,i}(L)f_{k,i}(L)\left(\frac{1}{\beta_{i}}-\frac{1}{\beta_{i}}\right)=0.
$$
Thus, from (\ref{orthapp}) and with $m_{n,i}\ne m_{k,i}$ we get (\ref{orthdelta}).

Now let us take the integral
\begin{equation}
m_{n,1}\int\limits_{-L}^{L}e^{3\sigma}\left(f_{n,1}(z)f_{k,2}(z)-{\tilde f}_{n,1}(z){\tilde f}_{k,2}(z)\right)dz.
\end{equation}
Performing the calculations, fully analogous to those presented in the previous paragraph, we get
\begin{eqnarray}\nonumber
(m_{n,1}+m_{k,2})\int\limits_{-L}^{L}e^{3\sigma}\left(f_{n,1}(z)f_{k,2}(z)-{\tilde f}_{n,1}(z){\tilde f}_{k,2}(z)\right)dz\\ \label{orthappmix}=-2e^{4\sigma(L)}\Bigl(f_{n,1}(L){\tilde f}_{k,2}(L)+{\tilde f}_{n,1}(L)f_{k,2}(L)\Bigr).
\end{eqnarray}
With (\ref{bounddeltaextra}), the r.h.s. of equation (\ref{orthappmix}) can be represented as
$$
-2e^{4\sigma(L)}\Bigl(f_{n,1}(L){\tilde f}_{k,2}(L)+{\tilde f}_{n,1}(L)f_{k,2}(L)\Bigr)=Ke^{4\sigma(L)}f_{n,1}(L)f_{k,2}(L)\left(\frac{1}{\beta_{1}}+\frac{1}{\beta_{2}}\right)=0.
$$
Thus, from (\ref{orthappmix}) we get (\ref{orthdeltamix}).

\section*{Appendix~E: The absence of tachyonic modes of the Higgs field}
The equation for the wave function of the $n$-th Kaluza-Klein mode, coming from (\ref{chieq}), takes the form
\begin{equation}
\left(e^{2\sigma}f_{\chi,n}'(z)\right)'+m_{\chi,n}^{2}f_{\chi,n}(z)-2\lambda_{br} v_{br}^{2}\delta(z-L)f_{\chi,n}(z)=0,
\end{equation}
where $m_{\chi,n}$ is the mass of this mode. Multiplying it by $f_{\chi,n}$ and integrating the result over the coordinate of the extra dimension, we get
\begin{equation}
m_{\chi,n}^{2}\int\limits_{-L}^{L}f_{\chi,n}^{2}(z)dz=\int\limits_{-L}^{L}e^{2\sigma}\left(f_{\chi,n}'(z)\right)^{2}dz+2\lambda_{br} v_{br}^{2}f_{\chi,n}^{2}(L).
\end{equation}
Since the r.h.s. of this equation and the integral in the l.h.s. are positive for $f_{\chi,n}(z)\not\equiv 0$, we get $m_{\chi,n}^2>0$.

Analogously, for equation (\ref{xieq}) we can obtain
\begin{equation}
m_{\rho_{i},n}^{2}\int\limits_{-L}^{L}f_{\rho_{i},n}^{2}(z)dz=\int\limits_{-L}^{L}e^{2\sigma}\left(f_{\rho_{i},n}'(z)\right)^{2}dz,
\end{equation}
leading to $m_{\rho_{i},n}^2\ge 0$.

\end{document}